# Reaction mechanisms of sodium carbonate activation in ultra-high-volume slag-cement blends

Samira Hossain[1,2,3], Zhanzhao Li[1,2,3], Kai Gong[1,2,3]*

[1] Department of Civil and Environmental Engineering, Rice University, Houston, TX 77005, United States

[2] Rice Advanced Materials Institute (RAMI), Rice University, Houston, TX 77005, United States

[3] Ken Kennedy Institute, Rice University, Houston, TX 77005, United States

* Correspondence author. Email: kg51@rice.edu

## ABSTRACT

Ultra-high-volume slag-cement (UHVS) blends can substantially reduce clinker ratio, but slow slag reaction limits strength development; $Na_2CO_3$ is a less caustic and lower-carbon activator for alkali-activated materials, yet its moderate alkalinity can similarly delay strength development. Here, dry-blended $Na_2CO_3$ (0–20 wt.% relative to water) was incorporated into a one-part binder containing 90 wt.% ground-granulated blast-furnace slag and 10 wt.% Portland cement (PC) to determine whether this integration could mitigate slow strength development in both systems. Compressive strength testing, isothermal calorimetry, time-resolved quantitative X-ray diffraction, Fourier-transform infrared spectroscopy, aqueous solution analysis, and thermodynamic modeling were combined to elucidate how $Na_2CO_3$ controls the evolution of phase assemblage, alkalinity, calculated porosity and strength development. At 5–10 wt.% $Na_2CO_3$ dosages, calcite-dominated carbonate reactions promoted a rapid alkalinity increase and accelerated early-age reaction and strength development; however, subsequent decline in alkalinity limited later-age slag reaction and gel formation, restricting continued strength development. In contrast, 20 wt.% $Na_2CO_3$ favored substantial early formation of gaylussite and carbonate-AFm phases, which moderated the initial pH rise and delayed early gel accumulation. The larger alkali inventory and evolving phase–solution partitioning helped sustain alkalinity at later ages, enabling continued slag dissolution and gel formation. This pathway produced higher reaction extent and lower calculated porosities with compressive strength development approaching that of neat PC. Screening-level cradle-to-gate life-cycle assessment indicates a 66–72% reduction in global warming potential relative to PC, depending on $Na_2CO_3$ production pathways. These results identify carbonate-mediated control of alkalinity persistence as key to sustaining slag reaction in UHVS systems.

**Keywords:** Ultra-high-volume blended cement; sodium carbonate activation; one-part AAM; quantitative X-ray diffraction; thermodynamic modeling; reaction kinetics and mechanism; life-cycle assessment

## 1 INTRODUCTION

Portland cement (PC) production accounts for approximately 8–9% of global anthropogenic $CO_2$ emissions, primarily because of limestone calcination and the high thermal energy demand of clinker manufacturing [1–4]. As global construction activity continues to expand [5,6], developing low-carbon cementitious binders that retain the engineering performance of conventional PC has become a central challenge in cement science [7–9].

Partial replacement of PC with supplementary cementitious materials (SCMs) is among the most effective strategies for reducing binder-related $CO_2$ emissions, as SCMs such as fly ash and ground granulated blast furnace slag (GGBS) generally have substantially lower embodied carbon than clinker [8–10]. However, prescriptive specifications and conventional practices typically limit SCM replacement to moderate levels, commonly ~25–35 wt.% for fly ash and other pozzolans [11,12] and ~25–50 wt.% for GGBS, although GGBS contents of up to 70 wt.% are permitted or used in some applications [13], depending on exposure conditions and performance requirements. While such blends reduce $CO_2$ emissions and heat of hydration and can improve durability, high replacement levels often slow early hydration, delay setting, and reduce early-age strength [14–17]. Other low-clinker systems, including limestone calcined clay cement ($LC^3$), can achieve ~40–50% clinker substitution while maintaining competitive later-age strength and durability in appropriately designed formulations [18–22]. However, early-age strength can remain lower, and further clinker reduction may compromise performance, requiring additional formulation or activation strategies [18]. GGBS is particularly attractive for further clinker reduction because of its latent hydraulic reactivity, yet conventional blended cements typically contain no more than 60–70 wt.% GGBS to ensure acceptable hydration kinetics under ambient conditions [16,23,24]. Transitioning into the ultra-high-volume range (>80 wt.%) generally results in protracted induction periods and insufficient early-age strength, motivating chemical activation, mineral additions, and processing strategies to overcome these kinetic barriers [24–29].

A related research direction involves alkali-activated slag (AAS), a class of clinker-free binders, in which dissolution of the glassy slag network releases Ca, Si, Al, and Mg species that subsequently form a dense matrix of alkali-containing calcium aluminosilicate hydrate (C–(N,K)–A–S–H) gel intermixed with secondary hydration products, such as hydrotalcite-like and AFm-type phases, depending on precursor and activator chemistry [30–33]. Under sufficiently alkaline conditions, slag dissolution and hydrate precipitation can proceed rapidly [30,34–36], yielding compressive strengths that approach or exceed those of PC and enhanced resistance to selected chemical exposures [37,38]. Despite the potential of alkali-activated systems, their production typically relies on highly alkaline activators such as NaOH and $Na_2SiO_3$, whose production can involve high embodied energy and associated emissions [9,39]. Depending on their composition and dosage, these activators may also cause rapid setting, high drying shrinkage [36,40,41] and other dimensional-instability concerns for AAS systems [42,43]. For instance, the drying shrinkage of $Na_2SiO_3$-activated slag can be several times greater than that of PC counterparts, presenting a significant hurdle for structural applications [40,44,45].

Sodium carbonate ($Na_2CO_3$) has emerged as a promising alternative activator because of its widespread industrial availability, lower causticity during handling, and potentially lower production-related embodied emissions than hydroxide- and silicate-based activators [46–49]. However, $Na_2CO_3$-activated slag systems typically exhibit slow reaction kinetics, extended induction periods, and delayed strength development [50–54], commonly attributed to the moderate initial alkalinity of $Na_2CO_3$ solutions, which is insufficient to drive rapid slag dissolution. The general hydration pathway, characterized by initial slow slag dissolution and carbonate precipitation, followed by a gradual increase in alkalinity and subsequent accelerated slag dissolution [46,50] and formation of C–(N)–A–S–H-type gel, is well established, although its kinetics depend strongly on activator dosage, calcium availability, and chemical composition of slag [46,50,51,55]. To accelerate the reaction and shorten the induction period, growing efforts have combined $Na_2CO_3$ with hydroxide- and/or silicate-based activators [54] or introduced Ca- or Mg-bearing additives such as CaO [56], $Ca(OH)_2$ [57], MgO [58], calcined layered double hydroxides [50], waste glass powder [59,60], and calcined recycled concrete fines [61].

These two research directions converge in ultra-high-volume slag-cement (UHVS) blends activated with $Na_2CO_3$, which occupy a distinct and relatively underexplored space between traditional PC–SCM blends and $Na_2CO_3$-activated slag binders. In these hybrid systems, we hypothesize that the small PC fraction hydrates more readily than GGBS under the initial reaction conditions, supplying Ca and contributing to the initial increase in alkalinity. The released $Ca^{2+}$ reacts with carbonate to form Ca-bearing carbonate phases. This increase in alkalinity promotes

GGBS dissolution and the subsequent formation of strength-giving hydrates. Such systems therefore offer a potential route to overcome the slow reaction kinetics and strength development of both UHVS blended cement and $Na_2CO_3$-activated slag, while retaining a one-part, dry-activated formulation.

Although a previous study examined $Na_2CO_3$ activation of UHVS blends containing up to approximately 90 wt.% GGBS, it primarily focused on macroscopic performance [62]. Consequently, the reaction mechanisms governing $Na_2CO_3$ activation in UHVS systems remain poorly resolved. In particular, the effects of $Na_2CO_3$ dosage on PC hydration, carbonate formation, alkalinity evolution, GGBS reaction, and binder-gel formation have not been quantitatively resolved over time. Resolving these coupled processes is necessary to understand how the initial reactions among PC, GGBS and $Na_2CO_3$ regulate subsequent reaction development and strength evolution in these hybrid binders.

In this study, these processes are investigated using a UHVS binder containing 90 wt.% GGBS and 10 wt.% PC as a model UHVS system, activated with varying $Na_2CO_3$ dosages. Hydration kinetics are quantified using isothermal conduction calorimetry (ICC), while phase assemblage evolution is examined through combined *in situ* and *ex situ* quantitative X-ray diffraction (QXRD) and Fourier-transform infrared spectroscopy (FTIR). Mechanical performance is evaluated through compressive strength measurements, while aqueous-chemistry trends in pore solutions are assessed using pH measurement and ICP-OES analyses. QXRD-derived reaction extents are then used as inputs for thermodynamic calculations of phase assemblage evolution, aqueous speciation, and porosity development, while the calculated aqueous-chemistry trends are compared with experimental measurements. By integrating these complementary analyses, this study elucidates how $Na_2CO_3$ dosage regulates reaction kinetics, phase evolution, and strength development, providing mechanistic insight into carbonate-mediated activation of UHVS blended binders and informing the development of one-part, low-clinker systems.

## 2 MATERIALS AND METHODS

### 2.1 Raw Materials and Mix Designs

Ground granulated blast-furnace slag (GGBS) was purchased from a commercial supplier, and Type I/II Portland cement (PC) was obtained from a local cement manufacturer (Sesco Cement Corporation, Houston, TX). The chemical compositions of both powders were measured by energy-dispersive X-ray fluorescence (EDXRF) using a Shimadzu EDX-8100 spectrometer equipped with a Rh anode X-ray tube (operated at 4–50 kV, 1–1000 µA, and air-cooled condition). Prior to measurement, powders were pelletized under a pressure of ~30 MPa for 2 mins using a hydraulic press (X-Press XP-400) to produce smooth, compact discs. The pelletized powders were analyzed under vacuum, and oxide concentrations were quantified using the fundamental parameter method. The measured bulk oxide compositions, summarized in Table 1, fall within the typical ranges reported for GGBS [63] and PC [64].

Powder densities were measured using $N_2$ gas pycnometry (BetterPyc380, Bettersize) and were found to be close to the typical ranges reported [65,66] (Table 1). Particle size distributions (PSDs) of the powders were measured using a laser diffraction analyzer equipped with a dynamic imaging system, wet dispersion unit, and ultrasonicator to mitigate particle agglomeration (BetterSizer S3 Plus). PSDs were calculated based on Mie scattering theory, assuming spherical particles, using refractive indices of 1.70 (real) and 0.1 (imaginary) for PC and 1.75 (real) and 0.10 (imaginary) for GGBS. The resulting PSDs are shown in Figure 1, with the corresponding $d_{10}$, $d_{50}$, and $d_{90}$ values summarized in Table 1. Based on the measured PSDs and powder density, the geometric surface area was calculated and is also given in Table 1. The results show that the GGBS used here exhibits finer particle sizes and a larger geometric surface area than the PC.

Table 1. Summary of bulk oxide compositions of ground granulated blast-furnace slag (GGBS) and Portland cement (PC) measured using X-ray fluorescence (XRF), along with powder density, particle size metrics ($d_{10}$, $d_{50}$, and $d_{90}$), and geometric surface area.

| **Properties** | | **Raw powders** | |
|---|---|---|---|
| | | **GGBS** | **PC** |
| Chemical compositions (wt.%) | CaO | 35.24 | 66.14 |
| | $SiO_2$ | 28.22 | 20.25 |
| | $Al_2O_3$ | 16.90 | 4.29 |
| | MgO | 10.64 | 1.44 |
| | $SO_3$ | 6.53 | 3.27 |
| | $Fe_2O_3$ | 0.52 | 3.54 |
| | $K_2O$ | 0.35 | 0.62 |
| | $TiO_2$ | 1.24 | 0.22 |
| | MnO | 0.24 | 0.09 |
| Physical properties | Density ($g/cm^3$) | 3.069 | 3.16 |
| | Specific surface area ($m^2/kg$) | 353.5 | 142.9 |
| | $d_{10}$ (μm) | 3.06 | 7.19 |
| | $d_{50}$ (μm) | 10.53 | 17.27 |
| | $d_{90}$ (μm) | 28.47 | 40.56 |

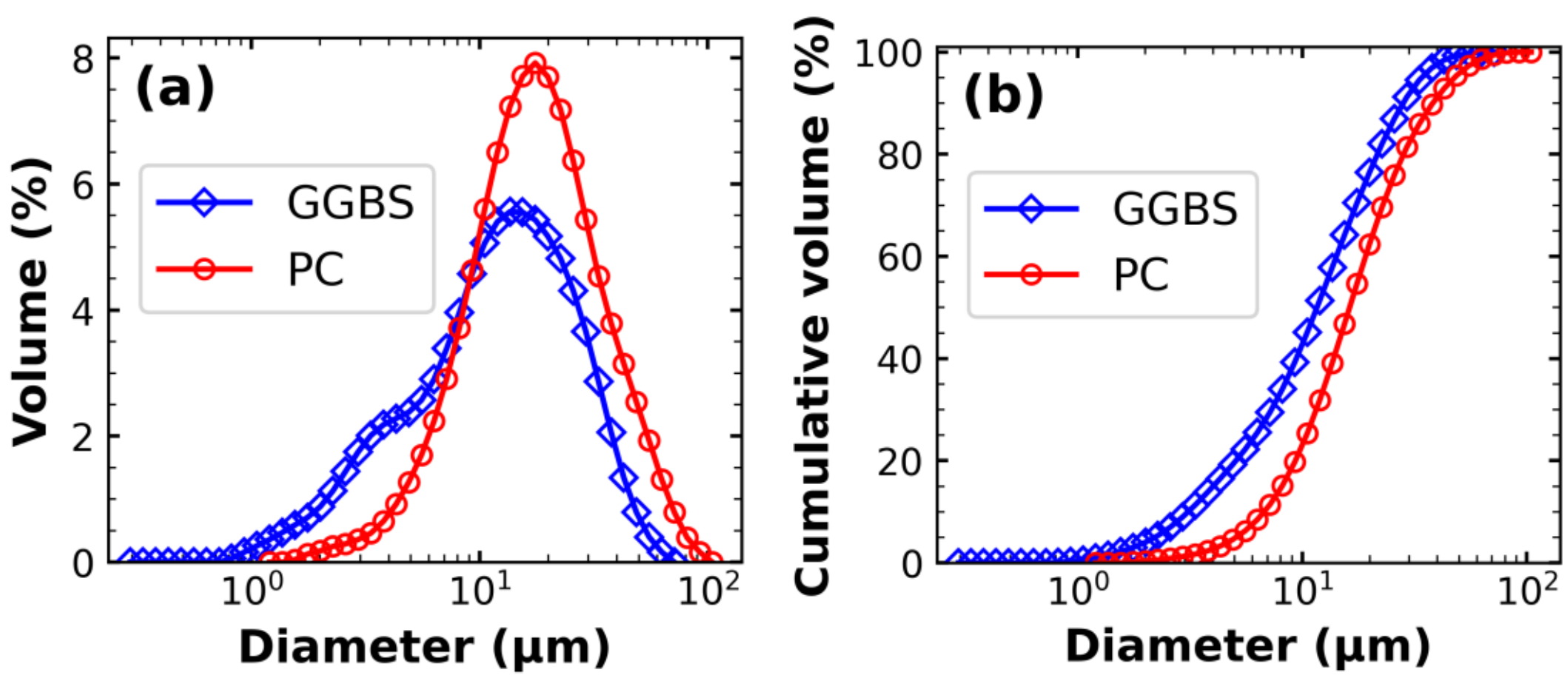

Figure 1. Particle size distributions of the raw GGBS and PC powders, as given by (a) differential volume fraction (%) and (b) cumulative volume fraction (%).

Overall, four UHVS blended cement formulations and one neat PC reference mix were investigated in this study, as summarized in Table 2. The blended cement binders consisted of 90 wt.% GGBS and 10 wt.% PC by mass, with a constant water-to-binder ratio (w/b = 0.6; binder = GGBS + PC) maintained across all formulations. Sodium carbonate ($Na_2CO_3$, 99.5% purity, Fisher Chemical) activator was introduced in powder form at dosages of 0 wt.% (denoted as 0SC), 5 wt.% (5SC), 10 wt.% (10SC), and 20 wt.% (20SC), expressed relative to the mass of mixing water. A neat PC paste (100PC) was prepared as a reference. Based on these mix designs, samples were prepared for isothermal conduction calorimetry (ICC), quantitative X-ray diffraction (QXRD), Fourier-transform infrared spectroscopy (FTIR), and compressive strength testing. Paste samples were used for ICC, QXRD, and FTIR analyses, whereas mortar specimens with a sand-to-binder ratio of 2 (binder = GGBS + PC) were prepared using cylindrical molds (diameter = ~21 mm, height = ~45 mm) for mechanical testing. Standard sand (ASTM-graded sand conforming to ASTM C778, supplied by U.S. Silica Company, Ottawa, Illinois, USA) was used for mortar preparation.

Table 2. Summary of mix designs for the UHVS blended cement systems and the neat PC reference, including the water-to-binder ratio and $Na_2CO_3$ dosage.

| No. | Mix ID | GGBS (wt.%) | PC (wt.%) | Water/(GGBS+PC) | $Na_2CO_3$/water (wt.%) | $Na_2CO_3$/ binder (%) | $Na_2O$eq/ binder (%) |
|---|---|---|---|---|---|---|---|
| 1 | 0SC | 90 | 10 | 0.6 | 0 | 0 | 0 |
| 2 | 5SC | 90 | 10 | 0.6 | 5 | 3 | 1.75 |
| 3 | 10SC | 90 | 10 | 0.6 | 10 | 6 | 3.51 |
| 4 | 20SC | 90 | 10 | 0.6 | 20 | 12 | 7.02 |
| 5 | 100PC | 0 | 100 | 0.6 | 0 | 0 | 0 |

## 2.2 Experimental Characterization

### *2.2.1 Isothermal conduction calorimetry*

ICC was performed on all mixes listed in Table 2 to investigate the effect of $Na_2CO_3$ dosage on the hydration kinetics of the UHVS systems. All measurements were conducted using an 8-channel TAM Air calorimeter (TA Instruments) at ~22 °C, following procedures consistent with our previous work [67]. For each test, the dry powders (GGBS, PC, and $Na_2CO_3$) were first homogenized by dry mixing and subsequently mixed with water using a vortex mixer for 2 mins. Immediately after mixing, 5 g of fresh paste was transferred into a standard glass ampoule, sealed, and placed in the calorimeter together with a reference ampoule containing 5 g of deionized water. Heat-flow and cumulative-heat data were recorded continuously for 14 days. The apparent overall degree of reaction of each mixture was estimated from the ICC data by normalizing the cumulative heat release to the estimated ultimate heat of reaction, following previously reported procedures [68,69]. Details of the calculation are provided in Section 1 of the Supporting Information.

### *2.2.2 In situ and ex situ quantitative X-ray diffraction*

The impact of $Na_2CO_3$ dosage on phase evolution in the UHVS systems (Table 2) was examined using both *in situ* and *ex situ* X-ray diffraction (XRD), followed by quantitative phase analysis. All XRD tests were conducted at room temperature using a Rigaku Smartlab II diffractometer operated in Bragg-Brentano geometry with Cu K$\alpha$ radiation at 40 mA and 40 kV. Scans were collected over a 2θ range of 5–60° with a step-size of 0.02° and a scan rate of 2°/min, corresponding to an acquisition time of ~30 mins per measurement. A high-purity corundum standard ($\alpha$-$Al_2O_3$, 99.99%, Beantown Chemical) was added to the dry powder as an internal reference prior to mixing with water to enable quantitative phase analysis. For the *in situ* measurements, paste samples were prepared following the same mixing procedure used for ICC, immediately loaded into a plastic sample holder and sealed with Kapton film (Figure S2 of Supporting Information) to minimize moisture loss and carbonation. After instrument calibration and background subtraction, *in situ* scans were collected continuously for ~12 hrs. The same specimens were then stored in a sealed container and analyzed *ex situ* at 1, 3, 8, 28, and 180 days using the same instrument settings and measurement conditions. Reference XRD patterns of raw GGBS, PC, corundum standard, deionized water, and the Kapton film were also collected under identical conditions. These reference patterns were used to determine the scattering parameters of the individual components, which formed the basis for the quantitative analysis described below.

### *2.2.3 Rietveld-PONKCS method for quantitative analysis*

Based on the XRD data collected, quantitative phase analysis was conducted using Rietveld refinement combined with the Partial or No Known Crystal Structure (PONKCS) method [70–76]. This approach enables quantification of poorly crystalline or amorphous phases, such as C–S–H-type gels and glassy phases in GGBSs and fly ashes, by treating them as calibrated pseudo-phases with empirically derived scattering factors. The method has been widely adopted for phase quantification in blended cements with SCMs [64,76,77] as well as alkali-activated materials [78–80]. A schematic overview of the refinement workflow adopted here is provided in Figure 2, and detailed refinement parameters are given in Section 3 of the Supporting Information.

Briefly, the process involved sequential calibration of structural factors and peak-shape parameters for individual components, which were subsequently fixed when analyzing the patterns for more complex hydrated paste samples. The corundum internal standard was first refined (Figure S3) to establish stable structure parameters, which were held constant in all subsequent refinements. The scattering contributions of Kapton film and free water were modeled with hkl phase using the Pawley method, following established literature procedures (see Figures S4 and S5) [72,74].

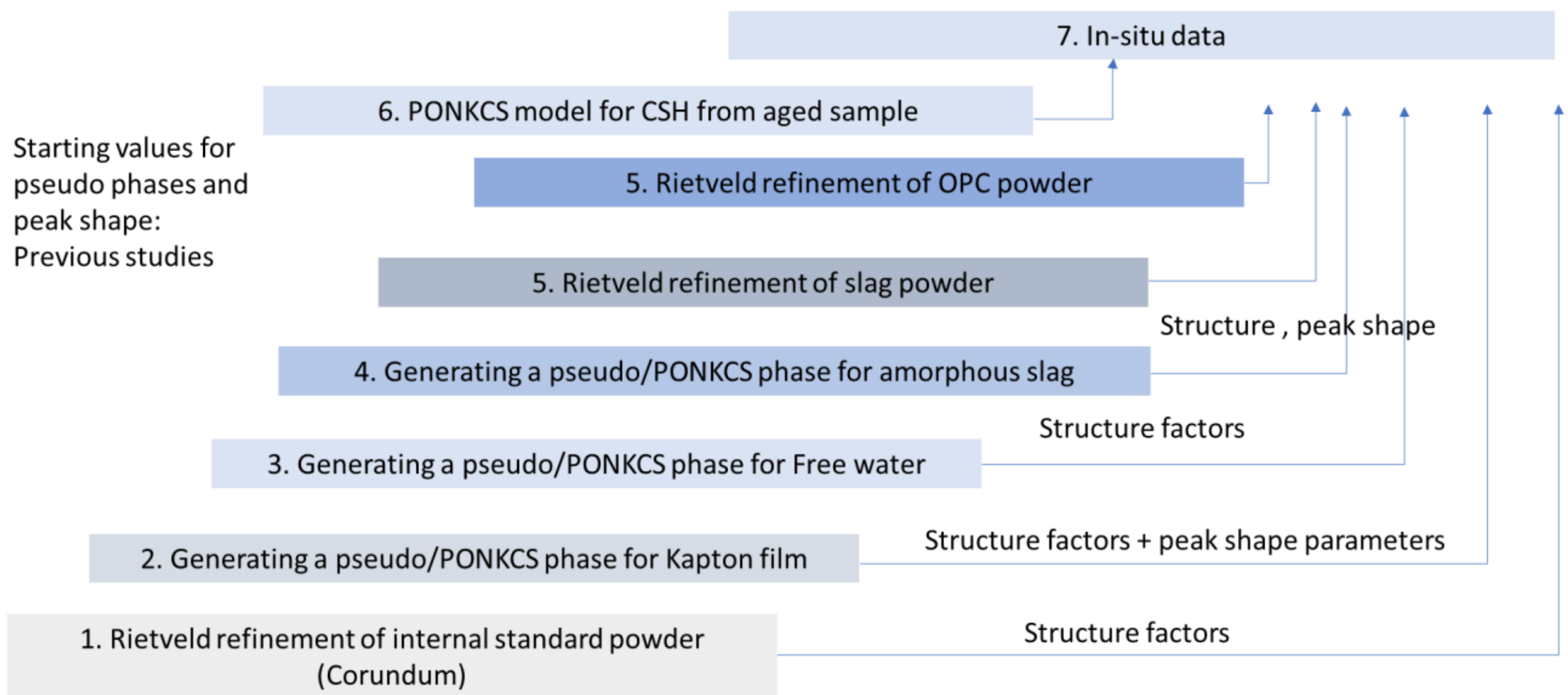

Figure 2. Schematic overview of the step-by-step Rietveld-PONKCS workflow used for phase quantification, developed based on several refs [71,76,81]. PONKCS = Partial or No Known Crystal Structure.

For the raw GGBS, which contains both amorphous and crystalline phases, the refinement proceeded in two stages. First, a high-order polynomial background was applied to model the amorphous hump, enabling the crystalline phases to be refined accurately without interference. After the crystalline phases were established, the background was redefined using a first-order Chebyshev polynomial, and the amorphous fraction was modeled using a PONKCS pseudo-phase. This pseudo-phase was generated by fitting the broad GGBS scattering hump to obtain the corresponding intensity and peak-shape parameters. This two-stage procedure yielded consistent structure factors and peak shape parameters for both the crystalline and amorphous fractions in the raw GGBS. A lower internal standard addition (10 wt.% corundum) was used for GGBS compared to PC (20 wt.%), following Stetsko et al. [71], to account for the much lower crystalline fraction in slag. For the predominantly crystalline PC powder, quantitative phase analysis was performed using conventional Rietveld refinement with 20 wt.% corundum as the internal standard. Refinement was carried out sequentially, beginning with background modeling using a first-order

Chebyshev polynomial (two parameters) with a 1/2θ term, followed by stepwise introduction of the dominant crystalline phases (e.g., corundum, alite, and belite) with refinement of their scale factors to ensure stable convergence. Unit-cell parameters were refined only when required and were constrained to vary within 1–3%, while peak-shape parameters and preferred-orientation corrections were adjusted as necessary.

Finally, for hydrated paste, a calibrated C–(A)–S–H pseudo-phase, modeled using the pseudo-Voigt peak set developed by Li and Scrivener [73], was included to account for the poorly crystalline C–(A)–S–H-type gel in 0SC and C–(N)–A–S–H-type gel in the $Na_2CO_3$-activated systems. Because the reference profile was derived from C–S–H formed in a different binder chemistry, the resulting quantity should be considered as PONKCS-assigned binding-gel fraction. The structural and peak-shape parameters obtained for individual phases from the above steps (Steps 1–6, Figure 2) were then fixed and applied to the analysis of both the *in situ* and *ex situ* XRD patterns. Phases were introduced sequentially from the most crystalline and abundant to the least crystalline and abundant, with background, unit-cell, peak-shape, and preferred-orientation parameters refined within constrained limits only when necessary. All quantitative analyses were performed using TOPAS version 7. Additional details of refinement procedure are given in Section 3 of the Supporting Information.

### *2.2.4 Fourier-transform infrared spectroscopy (FTIR)*

FTIR spectroscopy (Nicolet FTIR Infrared Microscope) was performed to monitor the impact of $Na_2CO_3$ dosage on the structural evolution of hydration products during early and later stages of reaction. Sample preparation followed the same procedure used for the ICC and XRD tests. Spectra were collected immediately after mixing, at 2-hour intervals during the first 12 hours, and subsequently at 1, 3, 8, and 28 days. Each spectrum was recorded over a wavenumber range of 500–4000 $cm^{-1}$ at a spectral resolution of 1.5 $cm^{-1}$ using 64 scans.

### *2.2.5 Compressive strength test*

Compressive strength tests were performed using a universal testing machine (Instron Model 5564, using 100 kN load cell) to evaluate the impact of $Na_2CO_3$ dosage on the mechanical properties of the UHVS mortar samples that have been sealed and cured for 1, 3, 8, and 28 days at room temperature (~22 °C). Cylindrical specimens (height-to-diameter ratio = ~2.2) were loaded under displacement-controlled conditions, and the maximum load at failure was used to calculate compressive strength. For each mixture and testing age, three replicate samples were tested, and the mean value and standard deviation were reported.

### *2.2.6 pH measurement of pore solution*

The evolution of pore solution alkalinity was monitored throughout hydration using a calibrated pH meter (VWR pHenomenal MU 6100L; SenTix ® 82 electrode). At early hydration ages (immediately after mixing to ~8 hrs), when the paste remained sufficiently fluid, the pore solution was separated by centrifugation and subsequently filtered through a 0.45 µm membrane filter prior to pH measurement. At later hydration ages (1 d, 3 d, and 8 days), when direct pore solution extraction was no longer feasible, an *ex situ* leaching method was employed following previously reported procedures [82,83]. Comparisons reported in the literature indicate that *ex situ* leaching can provide pH values reasonably consistent (± 0.1 pH unit for PC-based systems) with those obtained from directly extracted pore solutions under certain testing conditions, although the agreement depends on factors such as the liquid-to-solid ratio, leaching procedure, and pore-solution alkalinity [82]. The hydrated paste was crushed into powder, and a representative portion was mixed with deionized water at a liquid-to-solid mass ratio of 1:1 and homogenized for around ~5 mins using a vortex mixer. The resulting suspension was then filtered through a 0.45 µm membrane filter, and the pH of the resulting leachate was measured immediately using the calibrated pH meter.

### *2.2.7 Inductively coupled plasma optical emission spectroscopy (ICP-OES)*

The filtered solutions obtained from the pore solution extraction and *ex situ* leaching procedures were further analyzed by ICP-OES using an Avio® Max 550 ICP-OES instrument (PerkinElmer, USA) to determine the concentrations of dissolved species. Prior to analysis, the solutions were further filtered using a 0.25 µm membrane filter to remove any residual solid to avoid potential clogging of the nebulizer during testing and then were diluted using 2% $HNO_3$ solution to bring elemental concentrations within the calibration range of the instrument. The concentrations of Na, Ca, Si, and Al were quantified using external calibration standards. The experimentally measured aqueous concentrations were subsequently used to assess consistency with the pore-solution composition and temporal trends predicted by the GEMS thermodynamic calculations as detailed in Section 2.3.

### 2.3 Thermodynamic Modeling

Thermodynamic calculations were performed using the Gibbs free energy minimization software GEMS (version 3) [84,85] to simulate equilibrium phase assemblages and pore solution chemistry in the hydrated UHVS systems at different ages. The model inputs included the bulk oxide compositions of the raw GGBS and PC obtained by XRF, together with the reacted fractions of individual PC clinker phases and GGBS quantified from QXRD analysis. The calculations were carried out using the PSI/Nagra general thermodynamic database [86] and the Cemdata18 database [87] for cement hydrates. For mixes 0SC and 100PC, the C–S–H-type gel was represented by the CSHQ solid-solution model [88], while for $Na_2CO_3$-activated mixes (5SC, 10SC, 20SC), the CNASH_ss model [89] was used to account for lower Ca/Si ratios and the incorporation of Al and Na. The aqueous phase was modeled with the extended Debye–Hückel equation [84], assuming the aqueous phase is dominated by NaOH in the $Na_2CO_3$-activated mixes and by KOH in the other mixes. All equilibrium calculations were conducted at a temperature of 22 °C.

Because equilibrium calculations do not independently reproduce kinetically stabilized transient phases, experimentally observed phase constraints were imposed where necessary. In particular, the time-dependent gaylussite content in 20SC was prescribed using an empirical growth–decay function fitted to the QXRD results. The measured reaction extents were interpolated between experimental ages to provide continuous model inputs. Details of the interpolation functions, constrained phases, and sensitivity analyses are provided in Section 4 and 5 of the Supporting Information. Gel-pore volume was estimated by assigning fixed intrinsic gel-pore fractions to the modeled C–S–H and C–(N)–A–S–H phases, while capillary porosity was calculated as the difference between the total pore volume and the estimated gel-pore volume. The calculation procedure and assumed gel pore fractions are provided in Section 5 in the Supporting Information.

### 2.4 Life Cycle Assessment (LCA) and Cost Analysis

A scenario-based screening-level life cycle assessment (LCA) was conducted to quantify the environmental impacts of the $Na_2CO_3$-activated UHVS systems examined in this study. The analysis was performed using openLCA (version 2.5) [90] following an attributional LCA framework [91], which accounts for process-level environmental burdens while excluding market-mediated or consequential effects. Environmental impacts were evaluated using the ReCiPe Midpoint (H) method [92], with a focus on global warming potential (GWP) expressed as $CO_2$-equivalents per kilogram of dry binder, including GGBS, PC, and alkali activator. A cradle-to-gate system boundary was adopted, encompassing upstream core processes associated with the production of $Na_2CO_3$ and final blended systems. The system includes three major subsystems: (i) PC manufacturing, (ii) GGBS recovery and processing, and (iii) alkali activator production. A schematic overview of the system boundary is provided in Figure S14 of the Supporting Information.

The evaluated scenarios included a neat-PC reference, unactivated UHVS binder (0SC), and UHVS binder activated by 20 wt.% $Na_2CO_3$ considering three $Na_2CO_3$ production pathways: (i) recovery of natural soda ash from trona ore [93]; (ii) production via the Solvay process [94]; and (iii) an emerging $CO_2$ capture-integrated pathway, wherein $Na_2CO_3$ is synthesized by reacting NaOH with captured $CO_2$ [95–97]. An additional scenario using a renewable-intensive energy mix was also evaluated for the $CO_2$ capture-integrated pathway [98]. The cradle-to-gate system boundaries of all scenarios are shown in Figures S14–S18 of Supporting Information.

Life cycle inventory (LCI) data were compiled from open-access databases and supplemented with literature sources where necessary. Primary sources included the NIST Construction Materials Database [99] by the Federal LCA Commons and U.S. Electricity Baseline [100], and chemical manufacturing datasets from the National Renewable Energy Laboratory (NREL). For scenarios lacking comprehensive open datasets, particularly those involving $CO_2$-capture-integrated $Na_2CO_3$ production pathway, foreground inventories were obtained from peer-reviewed articles [95,96,101,102] and industry reports [94,103]. Detailed inventory assumptions, materials and energy flows, emissions factors, geographic and temporal coverage, and electricity-mix definitions are provided in Tables S4–S11 of Supporting Information.

A screening-level material cost analysis was conducted to compare the relative economic implications of the binder formulations and $Na_2CO_3$ production pathways. Material costs were calculated from the constituent mass requirements of each binder formulation and the corresponding unit prices. The adopted prices were intended to provide a comparative estimate rather than a detailed techno-economic assessment. Uncertainty in the calculated environmental and economic indicators was evaluated using Monte Carlo simulation [104]. More details on price assumptions, data sources, and uncertainty analysis are provided in Section 6 of Supporting Information.

# 3 RESULTS AND DISCUSSION

## 3.1 Strength Evolution of $Na_2CO_3$-activated Ultra-high-volume Slag-cement (UHVS) Blends

To evaluate if $Na_2CO_3$ activation could enhance the mechanical performance of UHVS blends, we first examined the compressive strength development of mortar samples prepared using the UHVS binders in Table 2. As shown in Figure 3, replacing 90 wt.% of PC with GGBS (100PC vs 0SC) significantly reduced compressive strength at all ages, with the 0SC mix reaching only ~1 MPa at 1 day and ~15 MPa at 28 days. This reduced strength development is consistent with prior reports on high-volume blended cement systems [24,105,106]. The addition of $Na_2CO_3$ enhanced early-age strength development of the UHVS blends at all studied dosages (5SC, 10SC, and 20SC) relative to the unactivated control mix (0SC). However, its effect on later-age strength (e.g., 28 days) depends strongly on $Na_2CO_3$ dosage. At lower dosage (5SC), the 28-day strength (~10 MPa) was even lower than that of the 0SC control despite its higher early-age strength. Increasing the $Na_2CO_3$ dosage to 10 wt.% and 20 wt.% resulted in significant gains across all ages. In particular, the 20SC mix achieved the highest 28-day strength (~33 MPa), which is more than twice that of 0SC and comparable to the neat PC reference. These results demonstrate that $Na_2CO_3$ dosage governs both early- and later-age strength development, with early acceleration at lower dosage not necessarily translating into improved later-age strength.

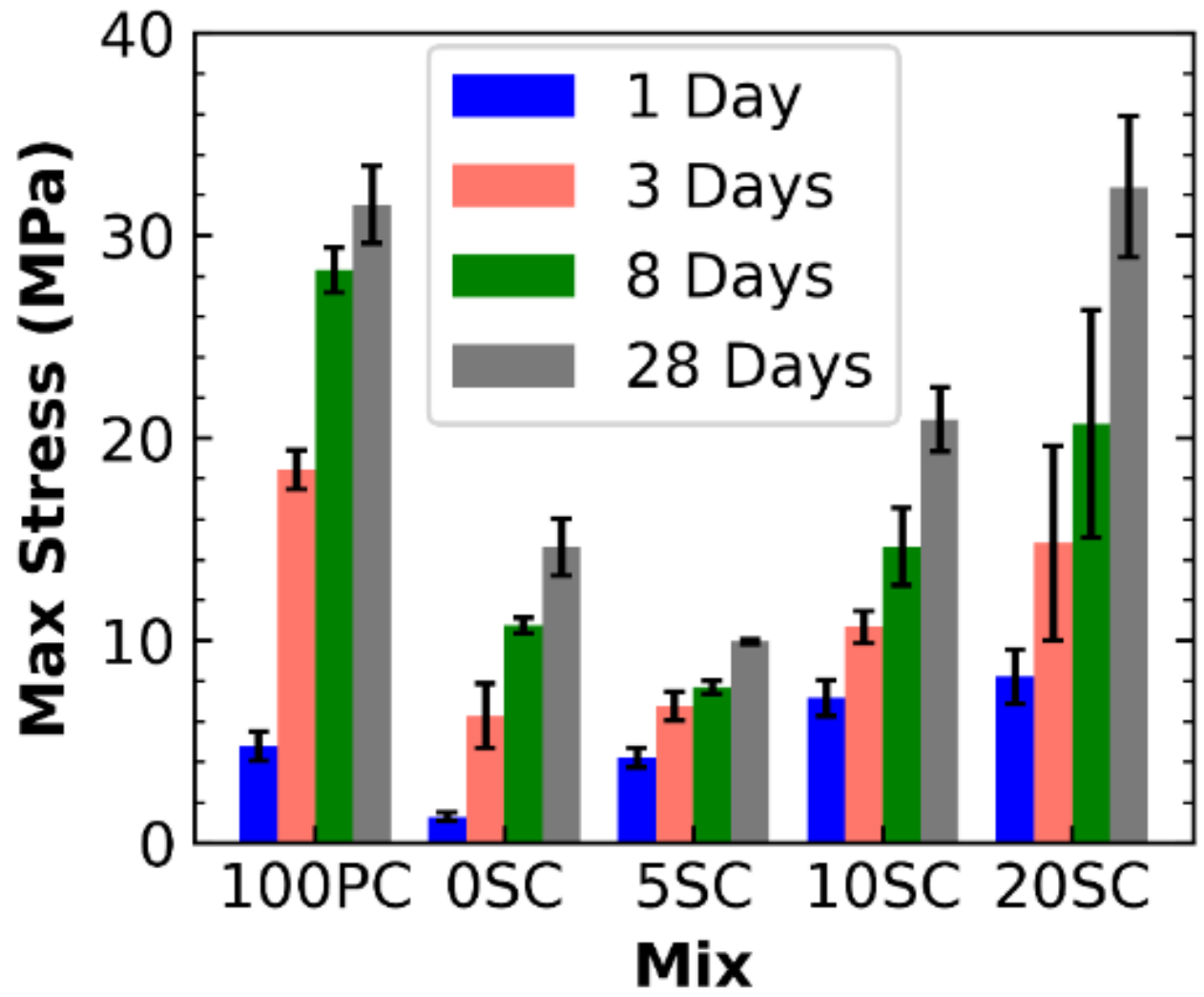

Figure 3. Compressive strength evolution of mortar specimens prepared with the five cementitious binders listed in Table 2. 100PC is the neat PC reference; 0SC, 5SC, 10SC, and 20SC denote 90% GGBS and 10% PC activated with 0%, 5%, 10% and 20% $Na_2CO_3$ (by mass of water), respectively. The values are based on the average of three specimens with the error bars denoting one standard deviation.

### 3.2 Hydration Kinetics of $Na_2CO_3$-activated UHVS Blends

To better understand the impact of $Na_2CO_3$ dosage on the mechanical properties of the UHVS blends, hydration kinetics were examined over 14 days using isothermal conduction calorimetry (ICC). Figure 4 presents (a) ICC heat flow and (b) cumulative heat curves for the five paste mixtures in Table 2, along with an additional reference paste containing only GGBS and water at a water-to-GGBS ratio of 0.6 ("GGBS"). All mixtures exhibited a sharp initial heat-flow peak immediately after mixing, which is commonly reported [46,50,67] and attributed to surface wetting and initial dissolution (see Figure S19 and more discussion in Section 7 of Supporting Information). Comparing the un-activated blend (0SC) with the neat-PC reference (100PC) shows that replacing 90 wt.% PC with GGBS delays the main hydration peak from ~8–10 hrs for 100PC to ~24 hrs for 0SC and substantially reduces its intensity (Figure 4a). Consequently, the cumulative heat released by 0SC is significantly lower than that of 100PC (~120 J/g vs. ~260 J/g at 250 h; Figure 4b). This behavior reflects the intrinsically lower hydraulic reactivity of GGBS relative to PC, as further evidenced by the lower cumulative heat release of the GGBS-only reference (~90 J/g at 250 h; Figure 4b) and explains the limited strength development observed for the 0SC mix (Figure 3). These trends are consistent with previous reports showing reduced hydration and strength development in blended cement systems with high-volume SCM [24,105].

The addition of $Na_2CO_3$ accelerates the hydration process, although the magnitude and persistence of the acceleration are strongly dependent on dosage. At 5 wt.% $Na_2CO_3$ (5SC), the main hydration peak shifts forward by ~9 hrs relative to 0SC, accompanied by a modest increase in peak intensity and early cumulative heat release (Figure 4). This early enhancement is consistent with the improved 1-day strength of 5SC (Figure 3). After ~3 days, however, the cumulative heat release of 5SC falls below that of 0SC, indicating that the initial acceleration was not sustained. This trend is consistent with the lower strengths of 5SC at 8 and 28 days (Figure 3). Increasing the $Na_2CO_3$ dosage to 10 wt.% (10SC) shifts the main peak further forward to ~3–4 h, even preceding that of 100PC (Figure 4a). This pronounced early acceleration corresponds to the higher 1-day strength of 10SC relative to 100PC. However, after ~1 day, the cumulative heat release of 10SC falls below that of 100PC, consistent with its lower later-age strength.

When further increasing the $Na_2CO_3$ dosage to 20 wt.% (20SC), the main reaction peak occurred at ~7–8 h, later than that of 10SC but close to that of 100PC (Figure 4a). The cumulative heat release of 20SC increases steadily, surpassing that of 10SC after ~10 hrs and remaining higher throughout the test period (Figure 4b). This more sustained heat evolution is consistent with the superior compressive strength of 20SC at all tested ages (Figure 3). Although 20SC has comparable 28-day strength to 100PC, its cumulative heat release remains noticeably lower after 1 day, reflecting the lower intrinsic hydration enthalpy of slag relative to PC (Section 1 in Supporting Information). Overall, cumulative heat release was observed to be positively correlated with compressive strength across all the mixtures studied, with linear regressions yielding $R^2$ values of 0.88–0.98 (Figure S20 in Section 7 of Supporting Information).

Furthermore, both 10SC and 20SC exhibit small heat-flow peaks preceding their main reaction peaks, indicating additional early-stage reactions. Prior studies on $Na_2CO_3$-activated slag systems containing Ca-rich additives, such as $Ca(OH)_2$, have reported the early formation of carbonate-bearing phases including calcium carbonate and gaylussite ($Na_2Ca(CO_3)_2 \cdot 5H_2O$) [46,55,107]. These phases may contribute to the early calorimetric features observed here. The origin and role of these features are examined through phase-evolution analysis in the following section.

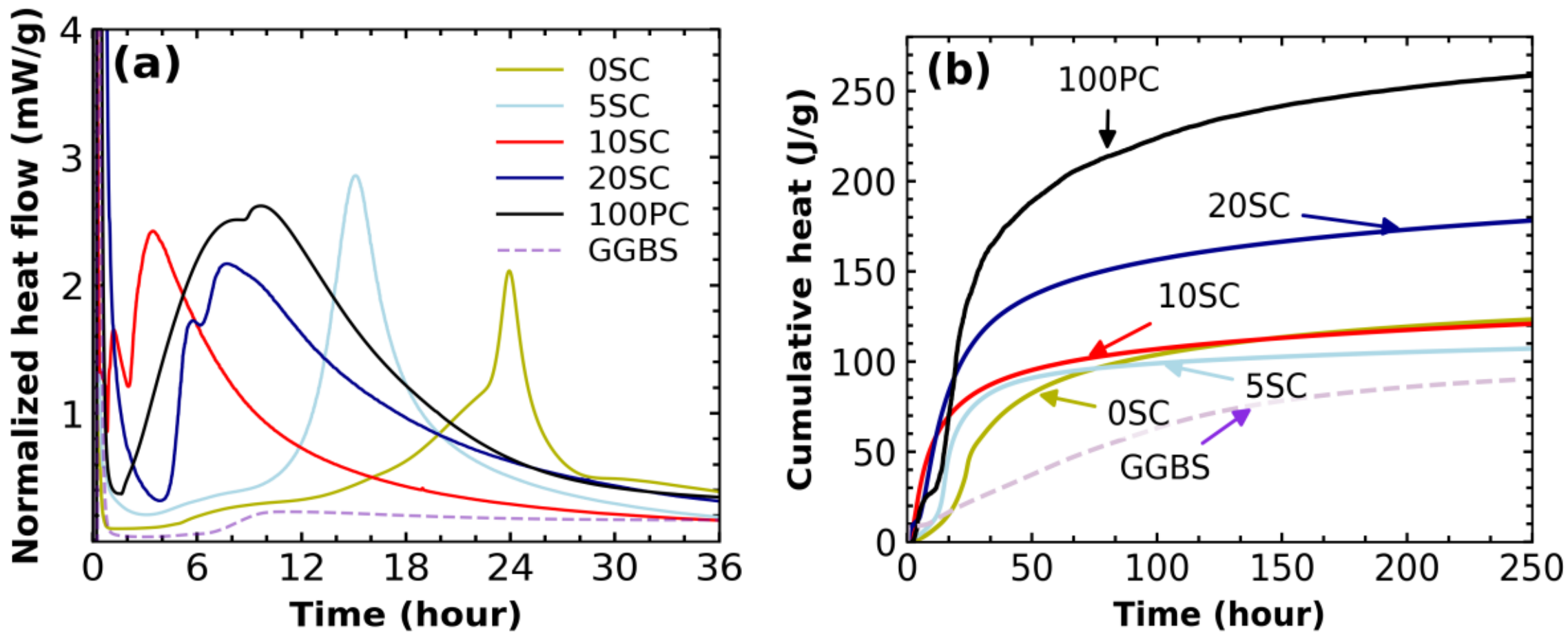

Figure 4. Isothermal conduction calorimetry (ICC) data for all studied binder systems: (a) heat flow and (b) cumulative heat release curves. In addition to the five binders listed in Table 2, a GGBS-only paste at the same water-to-binder ratio of 0.6 is included for comparison.

### 3.3 Phase Evolution during Hydration

#### *3.3.1 In situ and ex situ X-ray diffraction*

To elucidate the phase assemblages underlying the hydration kinetics described above, *in situ* and *ex situ* X-ray diffraction (XRD) was conducted on all mixes to monitor their phase evolution from the initial reaction period to extended curing ages. *In situ* measurements captured the initial ~12 hrs of hydration, while *ex situ* scans were collected on the same specimens at 1, 3, 8, 28, and 180 days. The corresponding diffraction patterns for the four slag–cement blends (0SC, 5SC, 10SC, and 20SC) are shown in Figures 5a–d, together with reference patterns of the raw GGBS, PC, and corundum internal standard. The raw GGBS pattern is dominated by a broad diffuse hump centered at ~30°, consistent with its predominantly glassy nature [108,109], along with minor crystalline phases such as gypsum and anhydrite. In contrast, the raw PC exhibits the expected crystalline phases, including alite, belite, aluminate, and ferrite. Corundum peaks are visible in both the raw powders and hydrated samples, as it was added as an internal standard for subsequent quantitative phase analysis.

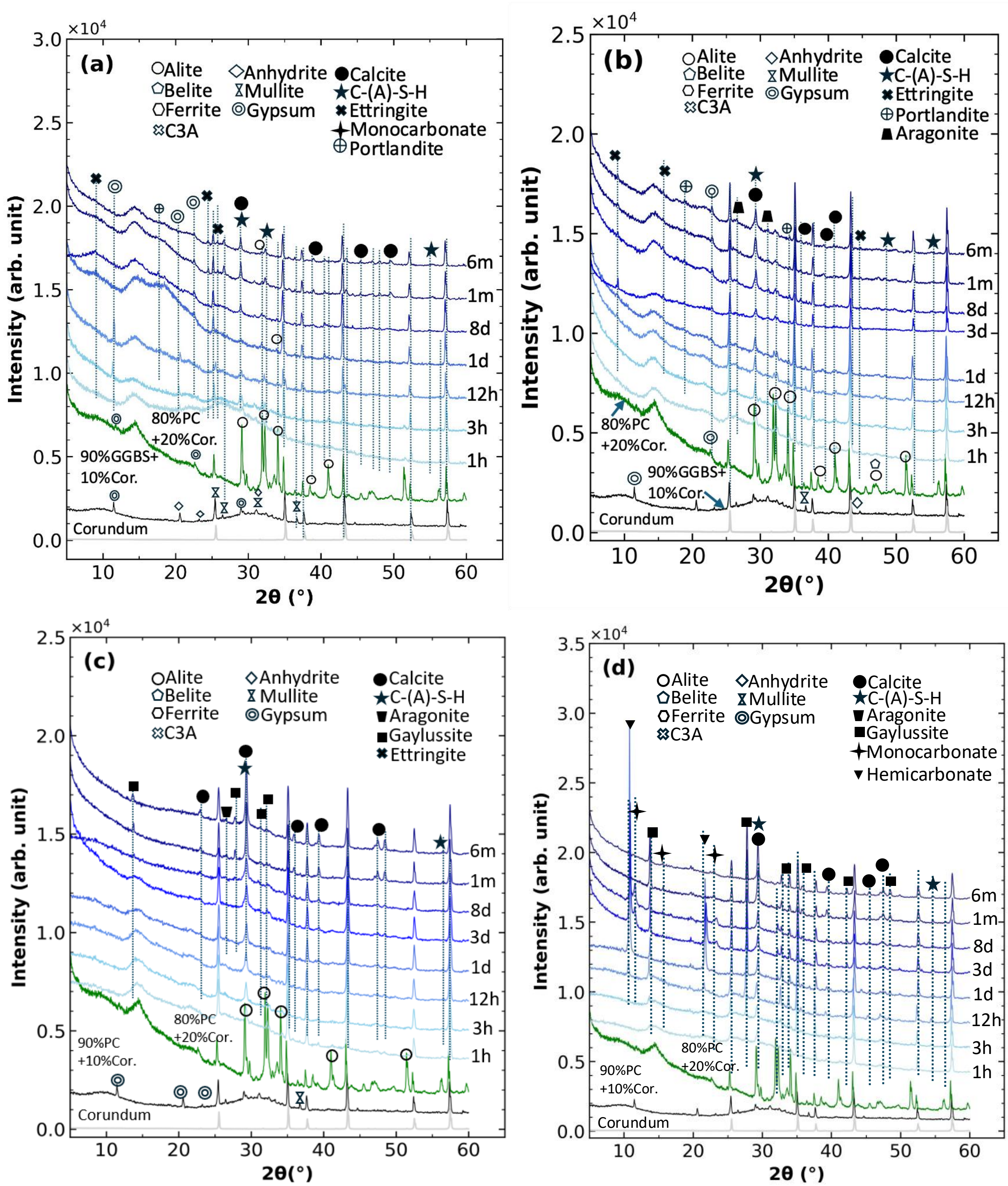

Figure 5. Evolution of X-ray diffraction (XRD) patterns as a function of hydration time (from 1 hour to 6 months) for the 90% GGBS and 10% PC mixes activated with (a) 0 wt.% $Na_2CO_3$ (0SC), (b) 5 wt.% $Na_2CO_3$ (5SC), (c) 10 wt.% $Na_2CO_3$ (10SC), and (d) 20 wt.% $Na_2CO_3$ (20SC) solution (by mass of water). XRD patterns of the raw materials, including the internal standard (corundum), 90% GGBS with 10% corundum, and 80% PC with 20% corundum, are also included for comparison. h = hour(s); d = day(s); m = month(s).

For the unactivated blend (0SC), the diffraction patterns are dominated by the corundum internal standard, with only modest changes over time, consistent with the low reactivity inferred from the calorimetry results (Figure 4). During the *in situ* scans over the first ~12 h, characteristic peaks of alite (~29.3°) from the PC fraction and gypsum from the raw GGBS and PC (~11.7° and ~20.7°) remain visible, reflecting limited early consumption of both phases. By 1 day, the gypsum reflections disappear along with emergence of ettringite peak at ~9.1°, consistent with the reaction of dissolved sulfate with $C_3A$ during early PC hydration [110–112]. After 1 day, the main alite peak at ~29.3° progressively decreases, while very weak portlandite reflections at ~18° emerge, consistent with early alite hydration and the delayed main calorimetric peak observed at ~24 hours. At later ages (> 8 days), the portlandite peak diminishes, reflecting its gradual consumption during slag reaction and potentially by carbonation. Meanwhile, the increasing intensity near 29.4° indicates gradual calcite formation due to carbonation.

The impact of $Na_2CO_3$ on the phase assemblage evolution is strongly dosage dependent. At 5 wt.% (5SC), the diffraction patterns (Figure 5b) show no major deviations from those of the unactivated 0SC system (Figure 5a), indicating that a lower carbonate dosage produces only modest changes in crystalline phase assemblage. Weak peaks assigned to aragonite and calcite become detectable at around 3 days. In contrast, higher $Na_2CO_3$ dosages lead to pronounced changes in phase evolution. At 10% (10SC), calcite reflections (e.g., those near 29.4° and 47.5°) appear within the first hour and intensify with time, while characteristic gaylussite peaks (22.6° and 13.5°) emerge after ~8 days, along with weak aragonite peaks.

At the highest $Na_2CO_3$ dosage studied (20SC), in contrast, gaylussite is detected within the first hour and its reflections increase rapidly thereafter, reflecting substantial early formation of Na-Ca carbonate hydrates under the higher carbonate dosage. Another distinctive feature of 20SC is the early formation of AFm-type carbonate phases. A weak reflection tentatively assigned to hemicarboaluminate (Hc) becomes visible at approximately 3 h, is clearly resolved by ~7 hrs and continues to increase thereafter. Monocarboaluminate (Mc) is subsequently detected at ~1–3 days. Both Hc and Mc fall below the XRD detection limit by ~28 days, indicating transformation, loss of long-range order, or redistribution of their constituent Ca, Al and carbonate as hydration proceeds because their stability is sensitive to pore-solution chemistry [113,114].

### *3.3.2 FTIR analysis*

While XRD effectively identifies crystalline phases, it provides limited sensitivity to poorly ordered hydration products, particularly C–S–H-type gels, whose main XRD reflections also overlap with calcite peaks. To complement the XRD results and better resolve those amorphous or poorly ordered components, FTIR analysis was performed. Figures 6a–d show the evolution of FTIR spectra over the range of 500–2000 $cm^{-1}$ for the 0SC, 5SC, 10SC, and 20SC systems, with full-range spectra provided in Figure S21 of Supporting Information. Spectra of the raw GGBS and PC powders are also included in each figure for reference. For the PC powder, the broad band centered at ~900 $cm^{-1}$ corresponds to Si–O asymmetric stretching vibrations in alite and belite [115–117]. Minor features at ~600, ~670, and ~1100 $cm^{-1}$ arise from $SO_4^{2-}$ bending and stretching modes associated with gypsum and anhydrite [118,119], consistent with the XRD results in Figure 5. Weak bands near ~872 and ~1432 $cm^{-1}$ correspond to the $\nu_2$ and $\nu_3$ modes of $CO_3^{2-}$, respectively, indicating the presence of calcite in the raw PC [115,117,120]. The raw GGBS spectrum is dominated by a broad band centered at ~900 $cm^{-1}$, characteristic of Si–O–T (T = tetrahedra Si or Al) asymmetric stretching in a depolymerized aluminosilicate glass network [31,119]. Minor sulfate-related features at ~600, ~670, and ~1100 $cm^{-1}$ are also observed, in agreement with the presence of gypsum and anhydrite in the XRD data.

For the unactivated blend (0SC; Figure 6a), a band at ~935 $cm^{-1}$ emerges at ~1 day and intensifies with time. This feature corresponds to the Si–O asymmetric stretching vibration of a C–S–H-type gel with relatively high Ca/Si ratios formed during PC hydration [121,122]. Its appearance is consistent with the main calorimetric peak occurring in near 1 day (Figure 4a). Sulfate-related bands at ~600 and ~670 $cm^{-1}$ persist during the first several hours but disappear by 1 day, in agreement with gypsum consumption and ettringite formation observed by XRD (Figure 5). All hydrated samples exhibit a pronounced band at ~1644 $cm^{-1}$, attributable to the H–O–H bending vibration of molecular water [117]. At 28 days, weak bands at ~873 and ~1417 $cm^{-1}$ indicate calcite formation [123,124], possibly due to mild sample carbonation. With the addition of 5 wt.% $Na_2CO_3$ (5SC; Figure 6b), the overall spectral evolution remains broadly similar to that of 0SC. The main difference is the appearance of a carbonate $\nu_3$ band near ~1410 $cm^{-1}$, together with a shoulder near ~1480 $cm^{-1}$. This shoulder is consistent with minor aragonite formation detected by XRD (Figure 5b). The Si–O asymmetric stretching band of the C–S–H-type gel shifts slightly to ~944 $cm^{-1}$, suggesting a marginal increase in silicate polymerization relative to 0SC [121,122].

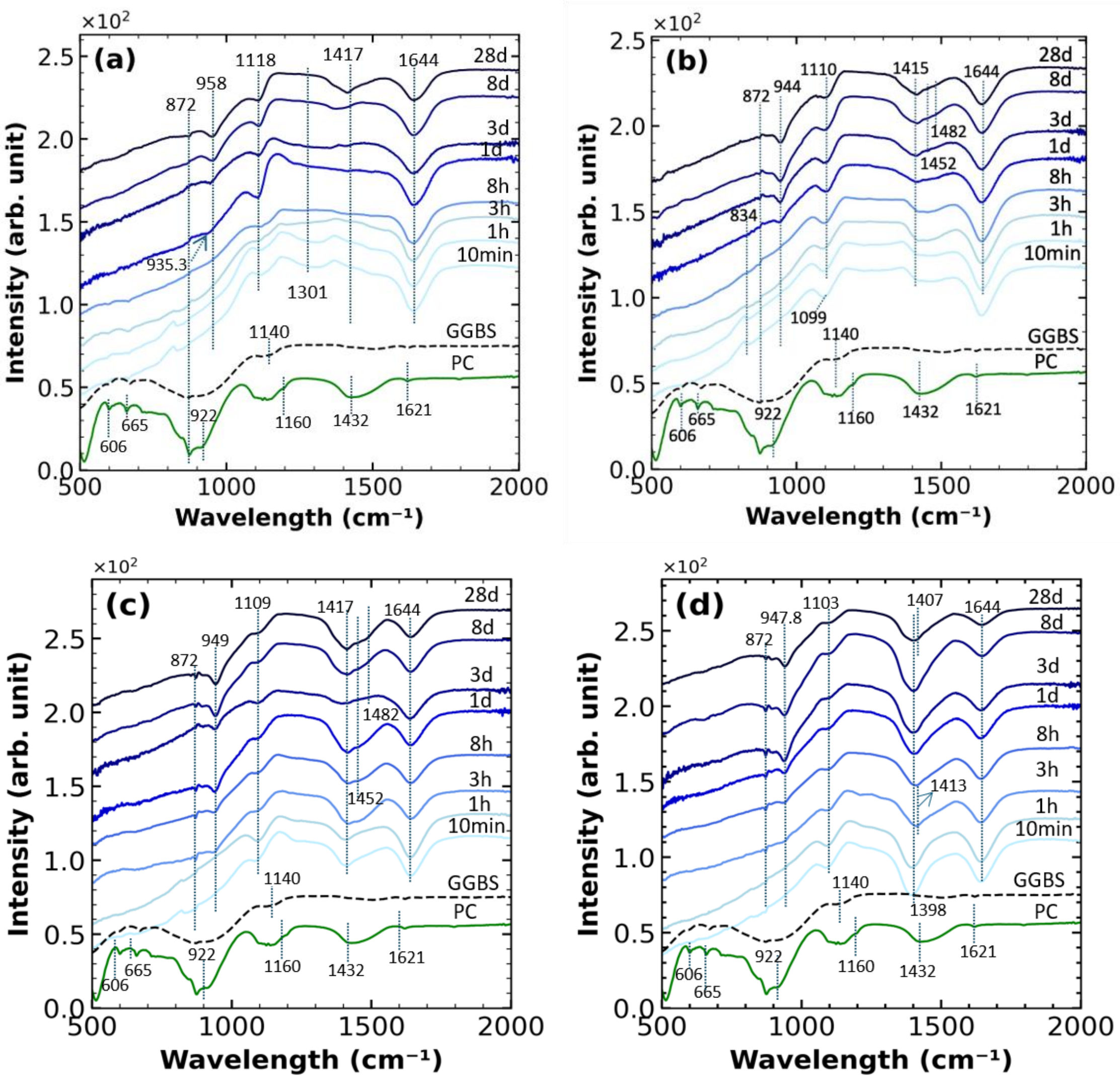

Figure 6. Evolution of FTIR spectra as a function of hydration time (from 1 hour to 28 days) for the 90% GGBS and 10% PC paste mixtures activated with (a) 0 wt.% $Na_2CO_3$, (b) 5 wt.% $Na_2CO_3$, (c) 10 wt.% $Na_2CO_3$, and (d) 20 wt.% $Na_2CO_3$ solution (by mass of water). Spectra of the raw GGBS and PC are also included for comparison.

With the addition of 10 wt.% $Na_2CO_3$ (10SC; Figure 6c), the C–S–H-related band (near ~944 $cm^{-1}$) is clearly observed at ~3 hrs and intensifies with time, consistent with the accelerated main hydration peak observed by ICC (Figure 4a). Carbonate bands at ~872 and ~1417 $cm^{-1}$ emerge after ~8 h, aligning with the XRD evidence for early and sustained calcite formation (Figure 5c).

A shoulder near ~1480 $cm^{-1}$ and a weak feature at ~856 $cm^{-1}$ are consistent with the presence of aragonite, corroborating the XRD results.

For the 20 wt.% $Na_2CO_3$ system (20SC; Figure 6d), the Si–O stretching band associated with C–S–H-type gel is also visible by ~3 hr, reflecting accelerated binder gel formation. The progressive shift of this band from ~938 to ~947 $cm^{-1}$ suggests increasing polymerization of the silicate network. This shift may arise from a decreasing Ca/Si ratio and/or increased Al incorporation into the silicate chains [122,125], both consistent with an increasing contribution of slag-derived species to the evolving gel in 20SC. Strong carbonate signatures are also evident: the calcite $\nu_2$ band at ~872 $cm^{-1}$ appears from ~3 hrs onward, while the broad $\nu_3$ carbonate-stretching band centered at ~1390–1420 $cm^{-1}$ may contain overlapping contributions from several carbonate-bearing phases such as calcite [124], gaylussite [123,126] and AFm-type carboaluminate [127] phases identified by XRD.

Overall, the FTIR results confirm that increasing $Na_2CO_3$ dosage promotes earlier C–S–H-type gel formation and produces more pronounced carbonate-phase development especially in 20SC.

### 3.4 Quantitative XRD Analysis Using the Rietveld-PONKCS Method

Although qualitative XRD and FTIR analyses reveal clear trends in phase evolution, quantitative analysis is necessary to evaluate how $Na_2CO_3$ dosage affects slag reaction and the formation of binder gels. Rietveld–PONKCS refinement was therefore performed on the *in situ* and *ex situ* XRD patterns using the internal standard method to quantify the evolution of phase assemblages and the extents of reaction over time. Details of the refinement workflow and the PONKCS model construction are provided in Section 3 of the Supporting Information.

#### *3.4.1 Refinement quality and method verification*

High-quality refinements were obtained for all the raw materials and hydrated samples. For the raw PC, Rietveld refinement yielded a weighted-profile residual $R_{wp}$ of ~7.1% (Figure 7a), with a quantified phase assemblage of ~61% $C_3S$, ~18% $C_2S$, ~5% $C_3A$, and ~12% $C_4AF$ (Table S1), consistent with compositions typically reported for Type I/II PC [77,128]. Back-calculated oxide compositions showed close agreement with those obtained from the XRF measurements (Figure S6; $R^2$ = ~1.00 for the line of parity), supporting the accuracy of the phase quantification. The raw

GGBS exhibits a high amorphous fraction (~97%) with an $R_{wp}$ of 7.5% (Figure 7b), consistent with the predominantly amorphous nature of GGBS [108,109]. Further verification using raw GGBS–PC blends with different nominal proportions showed close agreement between the nominal and QXRD-quantified GGBS contents (Figure 7c), further confirming the accuracy of the quantification approach. A representative refinement of the blended raw powder is provided in Figure S7 of the Supporting Information.

For hydrated samples, the broad scattering contribution from poorly crystalline C–S–H- or C–(N)–A–S–H-type gel was represented using a PONKCS pseudo-phase adapted from the C–S–H model developed by Li and Scrivener [73]. Because the reference profile was derived from C–S–H formed in a different binder chemistry, it was applied here as an approximate representation of the gel formed in the slag-rich cementitious binders. The effective ZMV parameter was independently calibrated using a year-old PC paste following the general PONKCS calibration approach described in [129]. The calibrated ZMV and pseudo-phase profile parameters were subsequently fixed for all hydrated samples for estimating the amount of evolving gel contribution.

Representative refinements for both *in situ* and *ex situ* hydrated samples are shown in Figures 7d-f and Figure S8 of Supporting Information. Across all mixtures and curing times, $R_{wp}$ values remain in the range of ~2.9–7.8% (Figure S9 of Supporting Information), below the threshold of 10% according to ref. [130]. This indicates good agreement between the measured and calculated diffraction patterns and supports the stability and reliability of the Rietveld–PONKCS refinement protocol across the investigated samples here.

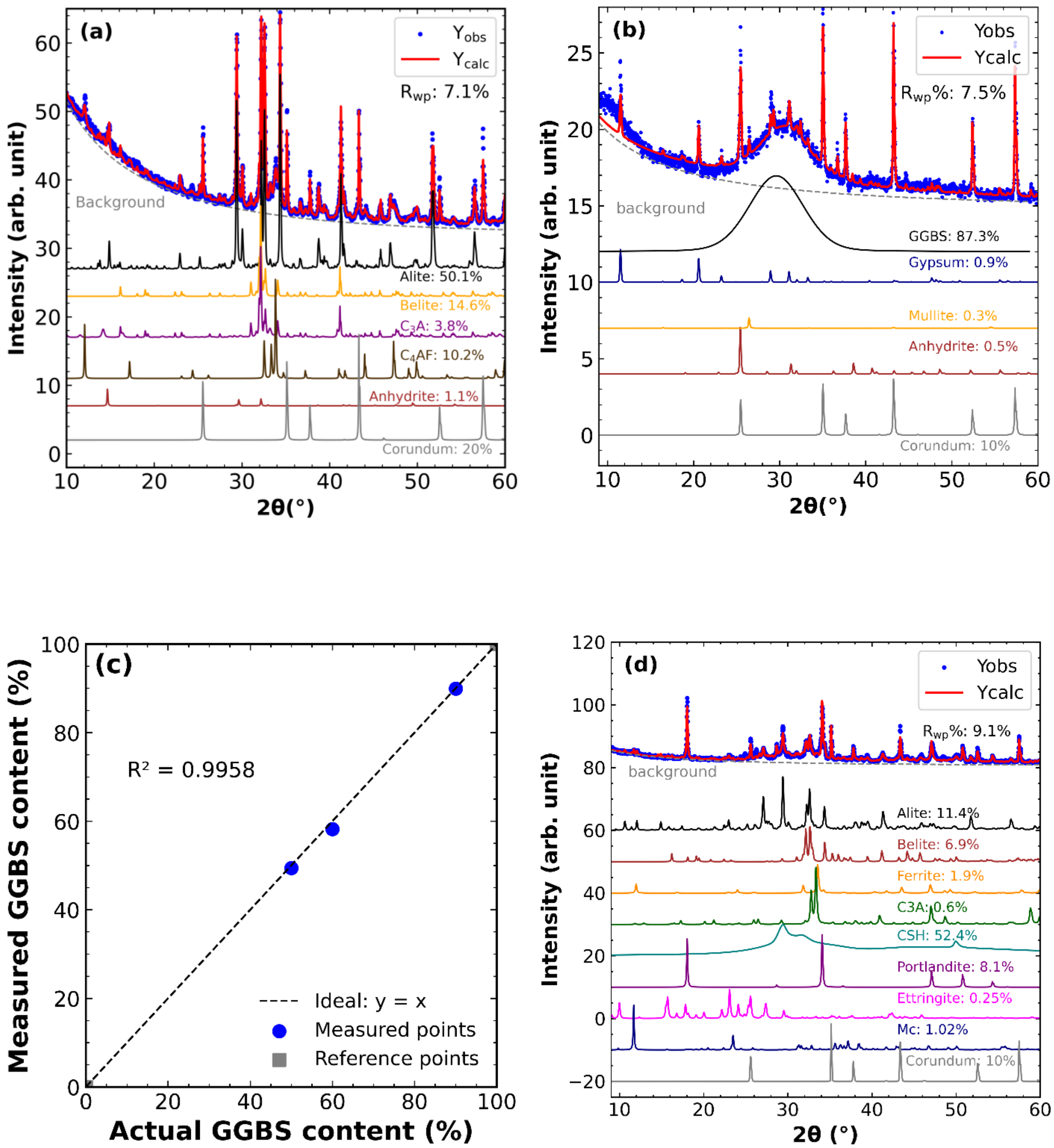
(a)
$Y_{obs}$
$Y_{calc}$
$R_{wp}$: 7.1%
Background
Alite: 50.1%
Belite: 14.6%
$C_3A$: 3.8%
$C_4AF$: 10.2%
Anhydrite: 1.1%
Corundum: 20%
Intensity (arb. unit)
2θ(°)
(b)
Yobs
Ycalc
$R_{wp}$%: 7.5%
background
GGBS: 87.3%
Gypsum: 0.9%
Mullite: 0.3%
Anhydrite: 0.5%
Corundum: 10%
Intensity (arb. unit)
2θ(°)
(c)
$R^2$ = 0.9958
Ideal: y = x
Measured points
Reference points
Measured GGBS content (%)
Actual GGBS content (%)
(d)
Yobs
Ycalc
$R_{wp}$%: 9.1%
background
Alite: 11.4%
Belite: 6.9%
Ferrite: 1.9%
C3A: 0.6%
CSH: 52.4%
Portlandite: 8.1%
Ettringite: 0.25%
Mc: 1.02%
Corundum: 10%
Intensity (arb. unit)
2θ (°)

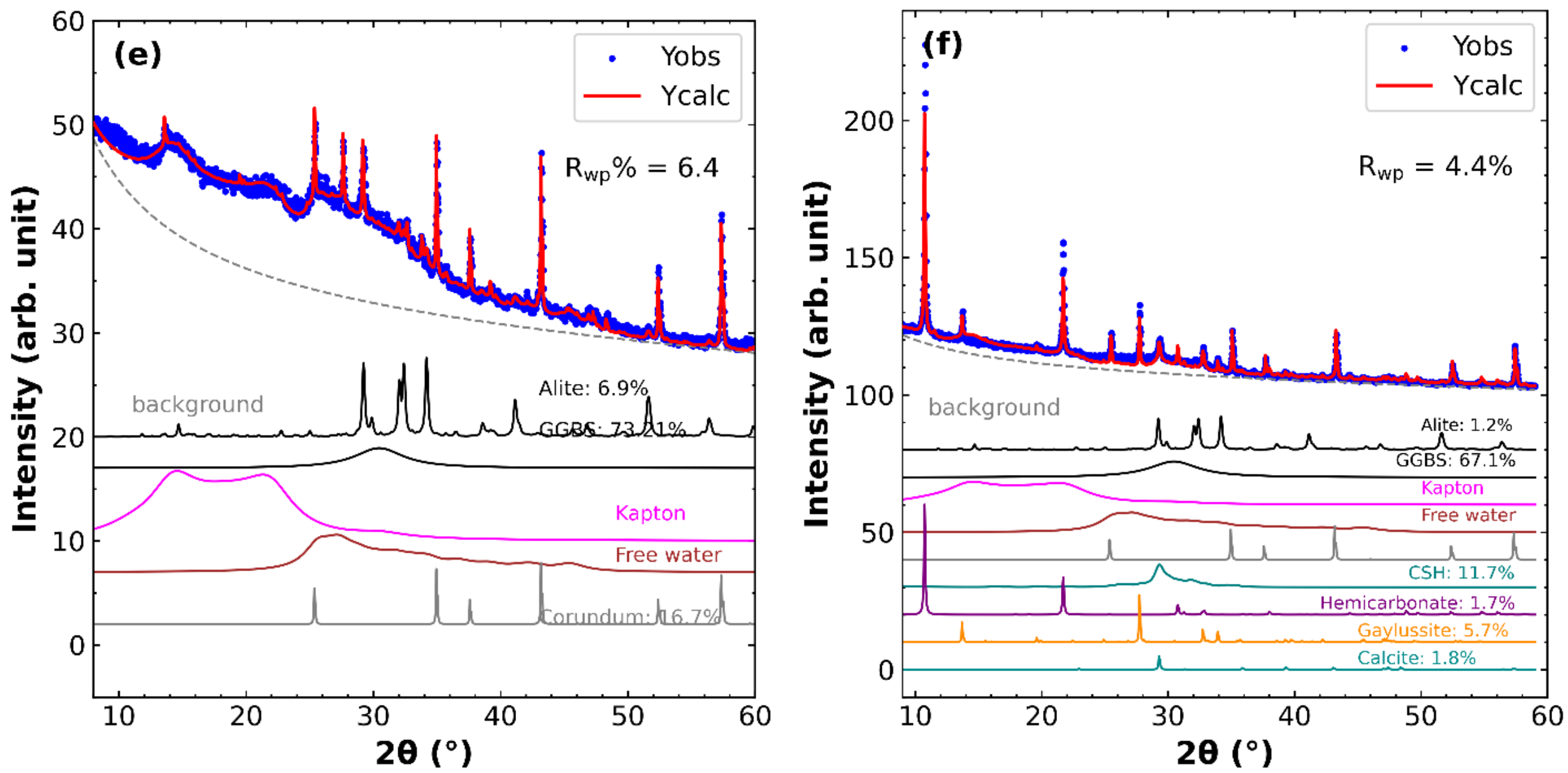

Figure 7. Representative quantitative XRD (QXRD) results to validate the Rietveld–PONKCS approach: (a) Rietveld refinement of raw PC powder; (b) QXRD refinement of raw GGBS; (c) comparison of nominal and QXRD-quantified GGBS contents in raw GGBS-PC blends with varying GGBS/PC ratios; (d) refinement of the 100PC paste hydrated for 13 months, used to calibrate the PONKCS gel pseudo-phase; and representative refinements of 20SC (10 wt.% PC-90 wt.% GGBS activated with 20 wt.% $Na_2CO_3$) at (e) ~30 mins of hydration (measured *in situ*), and (f) 8 days (measured *ex situ*). Additional refinement details and representative fits are provided in Section 3 and Figure S8 of the Supporting Information.

### *3.4.2 Effect of $Na_2CO_3$ dosage on phase evolution and reaction extent*

The QXRD results for the four UHVS blends are summarized in Figure 8, which shows the impact of $Na_2CO_3$ dosage on the time-resolved evolution of reactants (slag and main PC clinker phases) and reaction products (the main binder gel and secondary phases). Some minor phases typically below 1–2 wt.% are not included in Figure 8 due to their higher uncertainty but are summarized in Tables S20-S23 in Supporting Information. Overall, the results demonstrate that $Na_2CO_3$ dosage strongly influences both the kinetics and extent of slag reaction.

In the unactivated system (0SC; Figure 8a), alite consumption occurs mainly during the initial 3–12 hrs and contributes to the broad shoulder preceding the main calorimetric peak (Figure 4a), while no substantial slag consumption or C–(A)–S–H-type gel formation is detected during this *in situ* period. A measurable decrease in slag content and a concomitant increase in gel formation first appear at ~24 hr, coinciding with the main calorimetric peak. This temporal alignment indicates that the main ICC peak of 0SC is primarily associated with gel precipitation and slag reaction. Although weak portlandite reflections are visible in its diffraction pattern (Figure 5a), its refined content remains below the reliable quantitative resolution of the present QXRD analysis.

The addition of 5 wt.% $Na_2CO_3$ (5SC; Figure 8b) does not lead to substantial changes to the overall phase evolution relative to 0SC. Alite hydration is slightly accelerated during the *in situ* period, which explains the slightly earlier heat release and higher 1-day strength. However, the reaction extents of PC and slag and the accumulation of C–(A)–S–H-type gel remain lower than those in 0SC at later ages (≥3 days). This suppressed later-age reaction aligns well with the lower cumulative heat (Figure 4) and limited strength development of 5SC (Figure 3).

Increasing the $Na_2CO_3$ dosage to 10 wt.% (10SC) markedly accelerates the early reactions (Figure 8c). Approximately two-thirds of the alite reacts within the first hour, accompanied by immediate calcite precipitation, which explains the minor heat-flow peaks preceding the main calorimetric peak at ~3 hrs (Figure 4a). Slag consumption is observed shortly thereafter and proceeds continuously, with the slag fraction decreasing from its initial value of ~90% to ~74% by ~12 hrs. Concurrently, the PONKCS-assigned gel content increases rapidly from ~1–2% at ~3 hrs to ~20% at ~12 hrs, closely mirroring the decrease in slag content. Slag consumption continues at later ages, and the gel content reaches ~46% at 180 days. While this gel content is comparable to that of 0SC, the residual slag content in 10SC is lower at 180 days (~40 wt.% vs 46 wt.%). The resulting difference in phase assemblage reflects the higher content of secondary phases in 10SC, particularly calcite at 28 and 180 days. At these later ages, a small fraction of gaylussite (1–2 wt.%) is also observed, indicating continued redistribution of carbonate among the solid phases. The greater overall reaction extent of 10SC helps explain its higher later-age strength relative to 0SC.

At 20 wt.% $Na_2CO_3$ (20SC; Figure 8d), hydration is characterized by rapid early alite reaction, extensive carbonate formation, and sustained long-term slag reaction and gel formation. Approximately one-third of the alite reacts within the first ~1 hr, coinciding with the immediate appearance of calcite (~3%) and gaylussite (~3%), and the early heat release observed by calorimetry (Figure 4a). Slag consumption begins after ~2 h, with the residual slag fraction decreasing from its initial value of ~90% to ~84% at ~7 hrs and then more rapidly to ~68% by ~12 hrs. Over the same period, binder gel formation remains limited up to ~7 hrs (~1-2%), before increasing to ~8% at ~12 hrs. This value is substantially lower than the ~20% measured for 10SC at the same age. The rapid slag reaction and gel formation between ~7–12 hrs coincide with the main ICC peak centered at ~8 hrs and with a pronounced redistribution of carbonate phases. Gaylussite content increases to a maximum of ~10% at ~7–10 hrs and subsequently decreases, while hemicarboaluminate (Hc) increases from less than 1% at ~7 hrs to ~7% by ~12 hrs. With continued hydration, Hc decreases to ~3–4% between 1 and 8 days and becomes undetectable by 28 days, while monocarboaluminate (Mc) appears and increases, reaching ~7% by 8 days before also becoming undetectable by 28 days. In contrast, calcite content remains nearly constant (~3%) through 8 days, before increasing to ~7-9% between 28 and 180 days.

At later ages, slag reaction continues, with the residual slag fraction decreasing from ~68% at ~12 hrs to ~28% at 180 days, while the PONKCS-assigned gel content increases substantially from ~8% to ~57% over the same period. Compared with 10SC, 20SC exhibits greater slag consumption during the initial 12 hrs (~22% versus ~16% of initial mass of slag). However, this enhanced early slag consumption does not translate into greater gel formation (~8% versus ~20% within 12 hrs), but instead coincides with more extensive carbonate formation (~18% vs ~4%). In contrast, from ~12 hrs to 180 days, 20SC shows markedly greater additional binder gel accumulation than 10SC (~49% versus ~26%), together with slightly higher additional slag consumption over the same period (~40% versus ~34%). These differences are consistent with the higher cumulative heat release and superior later-age compressive strength observed for 20SC relative to 10SC.

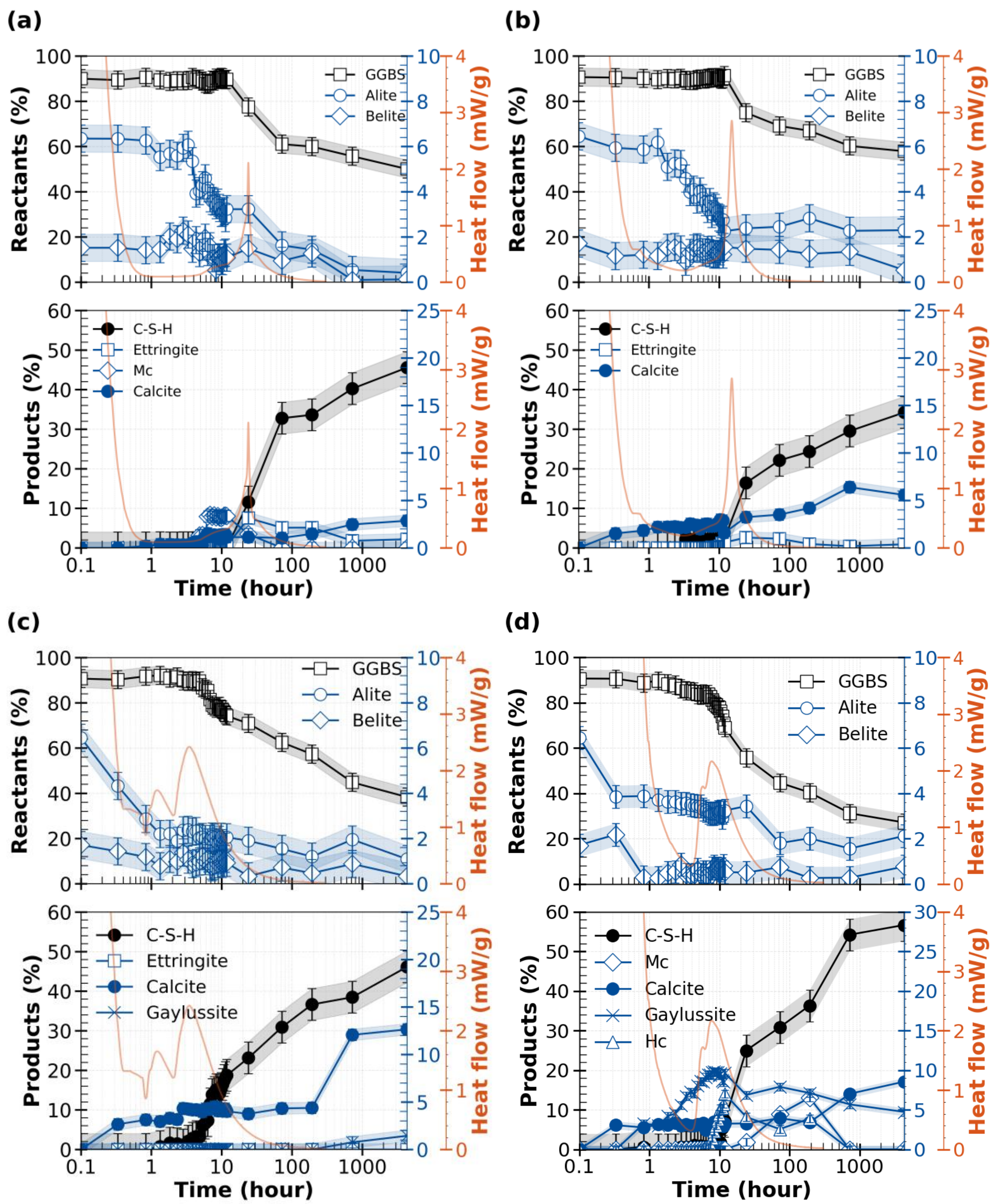

Figure 8. Time-resolved evolution of the major reactants and reaction products in weight percentage as quantified by QXRD using the Rietveld-PONKCS method for (a) 0SC, (b) 5SC, (c) 10SC, and (d) 20SC. The corresponding ICC heat-flow curves are overlaid for direct comparison.

The time-resolved overall reaction extent estimated from QXRD ($\alpha_{XRD}$) is compared with the corresponding ICC cumulative heat release for all mixtures in Figure 9a. A strong, approximately linear relationship is observed across all mixtures and curing ages, with an $R^2$ value of ~0.96 for linear regression. Despite substantial differences in binder composition and reaction pathways, the data for the neat-PC reference and the $Na_2CO_3$-activated slag–cement systems largely collapse onto a common trend, with no clear mixture-specific offsets or systematic deviations. Figure 9b compares the QXRD-derived degree of reaction ($\alpha_{XRD}$) with that estimated from ICC cumulative heat release ($\alpha_{ICC}$) using estimated ultimate reaction enthalpies of 479 $Jg^{-1}$ for PC (consistent with Ref. [131]) and 400 $Jg^{-1}$ for GGBS [132], following the method reported in [68,133] (see Section S1 of Supporting Information for calculation details and sensitivity analysis). Most data fall close to the line of equality, with an $R^2$ value of ~0.97 for linear regression. This close agreement demonstrates strong consistency between the calorimetric and QXRD-derived reaction trends and supports the reliability of the time-resolved QXRD analysis.

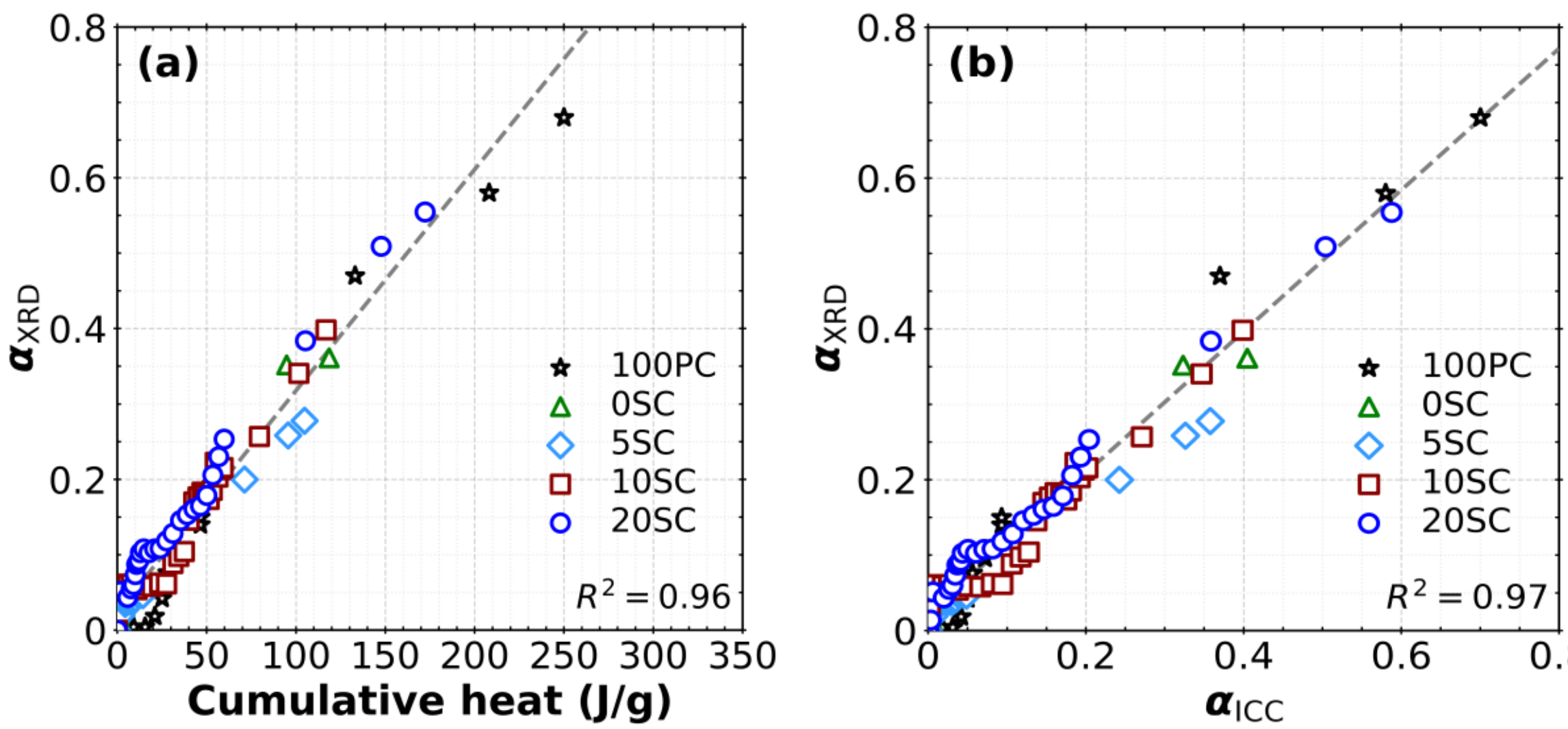

Figure 9. Correlation between the time-resolved overall degree of reaction quantified by QXRD and the corresponding (a) ICC cumulative heat release, and (b) extent of reaction estimated independently from ICC data (detailed calculation given in Section S1 of Supporting Information), for all mixtures and curing ages. The $R^2$ values represent goodness-of-fit from linear regressions.

The QXRD-derived degree of reaction also shows a strong positive correlation with compressive strength across all mixtures and ages studied, with an $R^2$ value of ~0.84 for linear regression (Figure 10). This relationship indicates that strength development is closely associated with the progression of reaction. Linear regressions for the individual mixtures yielded $R^2$ values of 0.91–0.97, although their slopes and apparent x-intercepts varied. The nonzero intercepts are consistent with the finite extent of reaction required to establish a mechanically connected microstructure in cementitious systems [134,135]. The differences in slope and intercept among mixtures further indicate that the relationship between reaction extent and strength is system-dependent, such that a given increase in reaction extent does not translate into the same strength gain across different mixtures. Together, these correlations support the consistency of the QXRD-derived reaction trends and demonstrate a clear link between reaction progress and strength across the investigated systems.

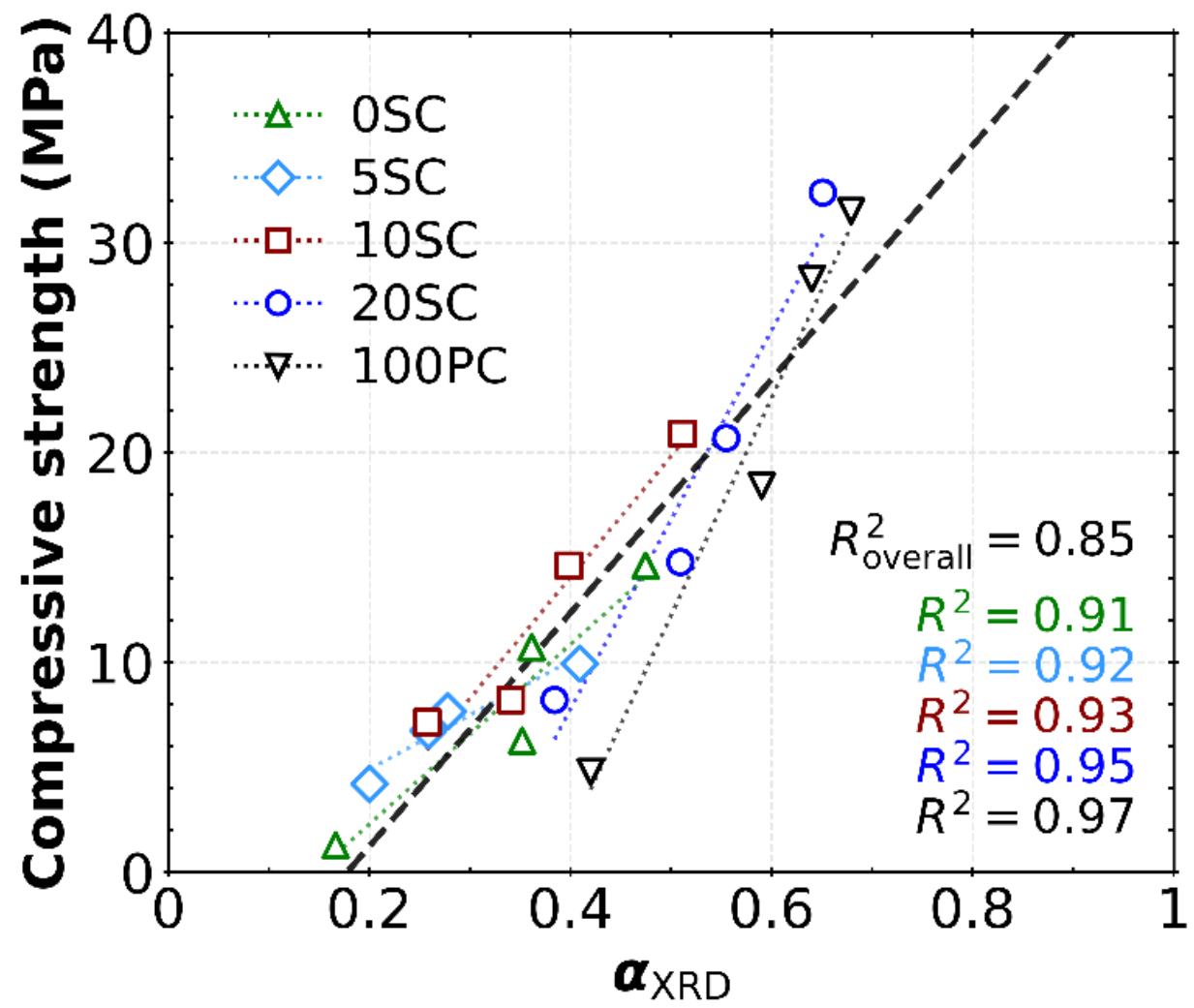

Figure 10. Relationship between compressive strength and the QXRD-derived degree of reaction for all mixes and curing ages. The $R^2$ value represents goodness-of-fit of linear regression.

### 3.5 Thermodynamic Modeling

Thermodynamic modeling was used to calculate equilibrium phase assemblages, pore-solution chemistry, and porosity at experimentally constrained reaction extents and to interpret the observed hydration and strength development of the $Na_2CO_3$-activated UHVS systems. At each experimentally investigated age, the QXRD-derived reaction extents of GGBS, alite, and belite were used directly as inputs to the equilibrium calculations. These reaction extents were additionally fitted using empirical functions to generate continuous, time-dependent inputs for the phase-evolution calculations and to evaluate the sensitivity of the modeled trends to interpolation. GGBS, alite, and belite were selected because their time-dependent consumption could be quantified with sufficient confidence from the QXRD results. The aluminate and ferrite each accounted for no more than ~2 wt.% of the total binder in the UHVS mixes, rendering their time-dependent consumption difficult to quantify reliably from the complex hydrated-paste patterns. Their full oxide inventories were therefore treated as available to the equilibrium assemblage at all modeled ages, effectively assuming complete reaction rather than imposing uncertain time-dependent reaction. Sensitivity analysis (Figure S13) showed that such treatment had little effect on the predicted hydration evolution and did not change the overall trends observed in the modeled phase assemblage and porosity. Similarly, calculations using empirically fitted rather than discrete QXRD-derived reaction extents produced no substantial changes in the overall thermodynamic trends (Section S5.2 and Figure S12).

An example of the fitting for 20SC is shown in Figure 11a. For this system, in addition to the interpolated reaction extents of the major reactants, the experimentally observed transient gaylussite content was represented using a growth-decay function (Figure 11b) and imposed as an additional time-dependent constraint in the thermodynamic calculations. This treatment was necessary because unconstrained equilibrium calculations cannot reproduce the kinetic persistence of metastable gaylussite, which constitutes a major transient sink for Na, Ca, and carbonate in 20SC and therefore influences the calculated aqueous chemistry. These empirical functions interpolate the experimentally measured reactant reaction extents and gaylussite evolution and should not be interpreted as predictive kinetic models. Further details of the fitting procedures and sensitivity analyses are provided in Sections 4 and 5 of the Supporting Information.

The resulting equilibrium phase evolution for 20SC is shown in Figure 11c, while corresponding results for the other mixtures are provided in Figure S11 of the Supporting Information. Across all investigated systems, the calculated assemblages included C–S–H or C–(N)–A–S–H-type gels, strätlingite, ettringite, and hydrotalcite-like phases. With the addition of $Na_2CO_3$, carbonate-bearing phases, including calcite and Mc, are also predicted to form. Because the equilibrium framework does not explicitly account for kinetic persistence, experimentally observed transient phases such as Hc are not reproduced unless explicitly imposed (as in the case of gaylussite). Such discrepancies are commonly reported in thermodynamic modeling of cementitious systems [33], and may arise from kinetic stabilization, evolving aqueous chemistry, and limitations in the available phase models and thermodynamic database. The calculated amounts of strätlingite and Mg–Al layered double hydroxide phases exceed those resolved experimentally. Similarly, model-experiment discrepancies have been reported for alkali-activated slags [33,136–138], and likely reflect limitations inherent to both experimental phase quantification and equilibrium thermodynamic modeling. Minor or poorly crystalline phases can be difficult to quantify accurately by QXRD, whereas equilibrium calculations may not capture kinetically persistent or metastable phases and are constrained by the available phase models and thermodynamic database.

Despite these limitations, the predicted quantity of the primary binding phase, i.e., C–S–H or C–(N)–A–S–H-type gels, shows good agreement with the QXRD-derived gel contents. As shown in Figure 11d, a strong linear correlation ($R^2$ value >0.94) is observed between thermodynamically predicted and experimentally quantified gel contents across all mixtures and curing ages. The larger deviations observed for 20SC largely reflect the discrepancies in its calculated secondary phases (e.g., Hc, strätlingite and hydrotalcite). Because the QXRD-derived GGBS, alite and belite reaction extents were used as model inputs, this comparison represents an internal consistency check rather than an independent validation. Nevertheless, agreement in the approximate magnitude and temporal evolution of the gel content, along with the sensitivity analyses, indicates that the model captures the principal equilibrium trends required to interpret associated pore-solution chemistry and porosity.

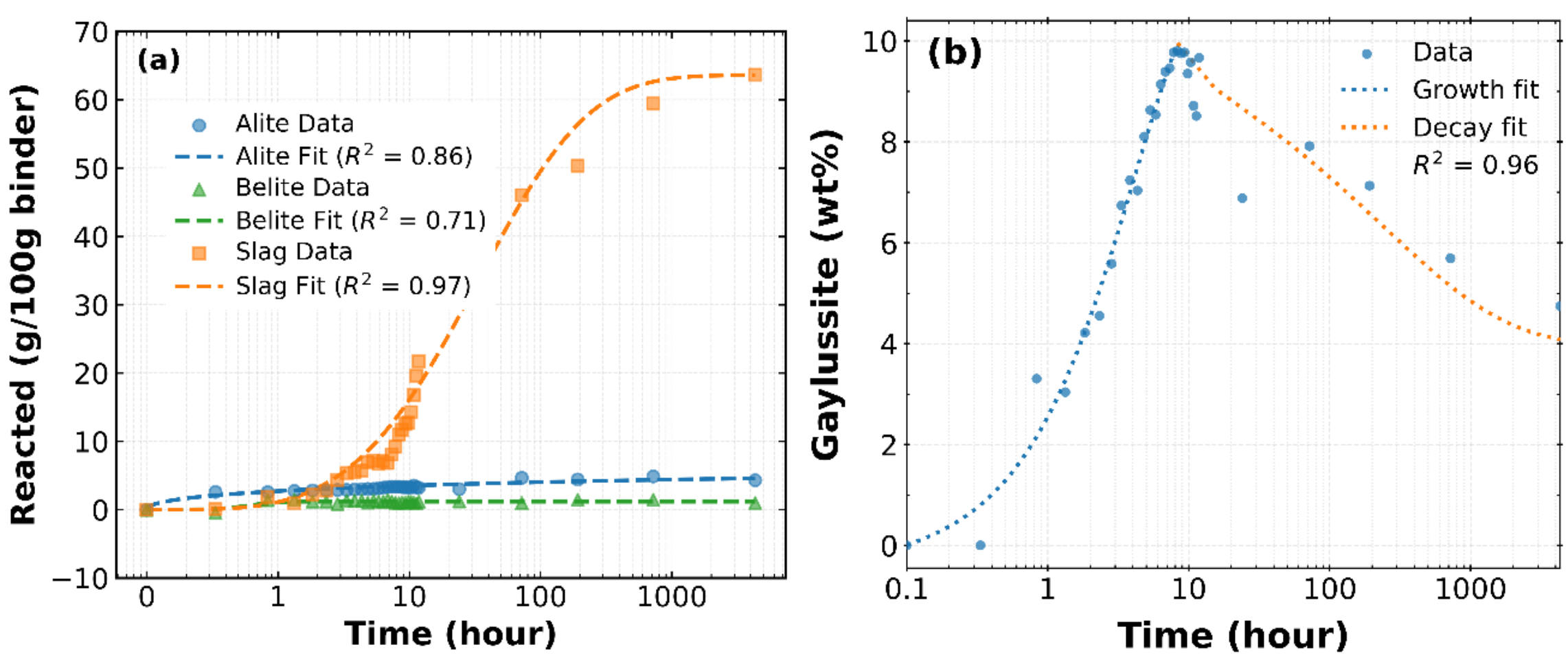

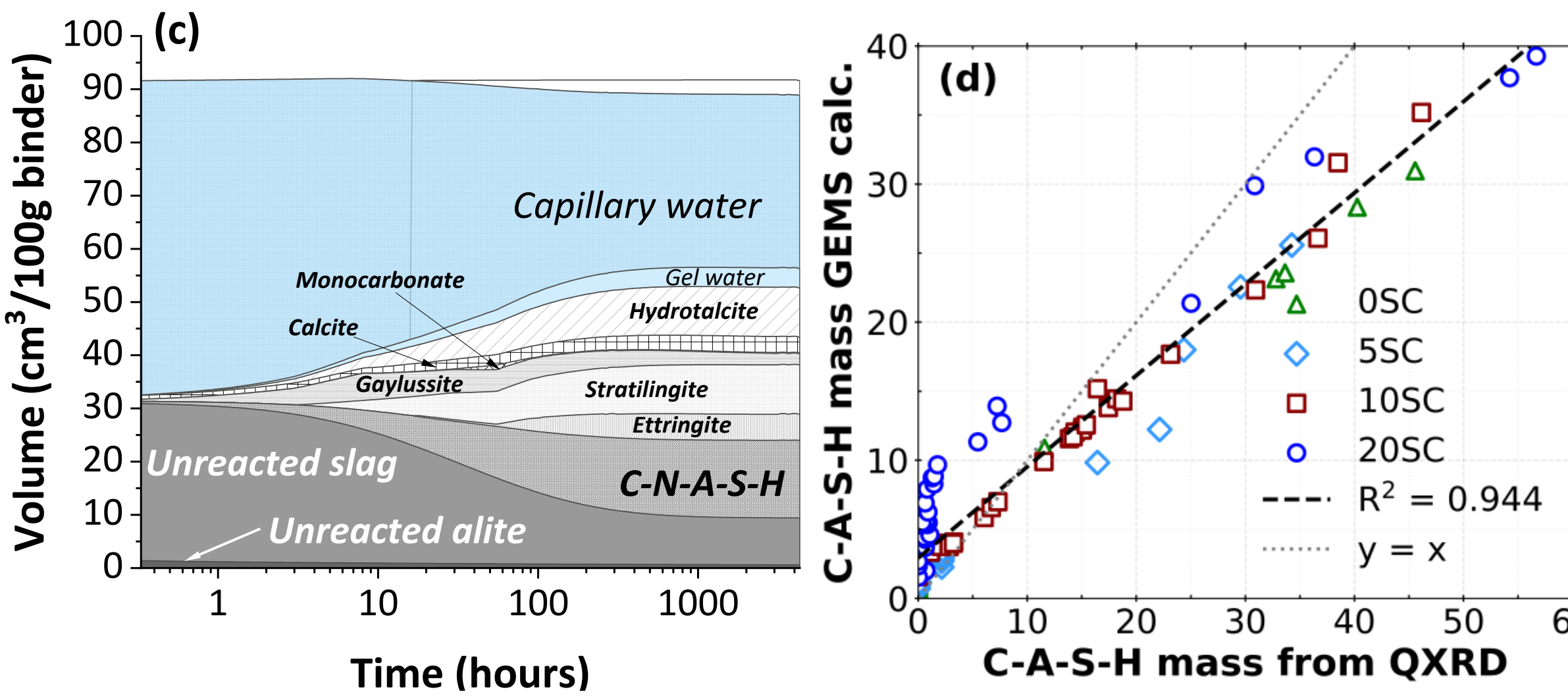

Figure 11. (a) Empirical sigmoidal fit used to interpolate the QXRD-derived reaction extents of major reactants in 20SC for input to the thermodynamic calculations in GEMS; (b) growth–decay fit representing the experimentally quantified transient gaylussite evolution imposed as an additional time-dependent constraint in GEMS; (c) GEMS-calculated phase assemblage evolution of 20SC based on the experimentally constrained inputs shown in (a) and (b); and (d) comparison of GEMS-calculated and QXRD-quantified gel contents for all mixtures and curing ages.

### *3.5.1 Evolution of pore solution chemistry*

Figure 12a compares the thermodynamically calculated evolution of pore-solution pH among the investigated binder systems. In the unactivated blend (0SC), the calculated pH rises rapidly to ~12.5 within the first hour due to dissolution and hydration of the PC fraction and subsequently stabilizes at ~12.2–12.3. The addition of $Na_2CO_3$ substantially alters both the magnitude and persistence of alkalinity. In 5SC, the calculated pH increases to ~13.5 within ~10 hrs but declines progressively at later ages. In 10SC, the pH rises more rapidly and reaches a maximum of ~13.8 within ~2 hrs, followed by a gradual decrease. In 20SC, the initial pH increase is lower than that of 10SC, but the calculated curves cross at approximately 12 hrs, after which 20SC maintains a higher pH of ~13.6 over the extended hydration period studied. Thus, $Na_2CO_3$ dosage affects not only the initial pH increase but, more importantly, the persistence of alkalinity.

The experimentally measured pH evolution (Figure 12b) qualitatively reproduces these key trends. The early-age values were measured in directly extracted pore solutions, whereas the later-age values were measured from *ex situ* leachates and therefore represent approximate alkalinity rather than *in situ* pore-solution pH. A direct comparison of the calculated and experimentally measured pH values is provided in Figure S22, showing broad trend-level consistency despite large differences in absolute values. Both approaches indicate substantially higher alkalinity in $Na_2CO_3$-activated systems than in 0SC, and importantly, stronger initial alkalinity in 10SC but more persistent alkalinity in 20SC, although the crossover occurs at different times (~12 hrs in calculations and ~24 hrs in experiments). Together these results indicate that sustained alkalinity, rather than the highest initial pH alone, enables prolonged slag reaction and gel formation in 20SC, consistent with the QXRD results in Section 3.4.

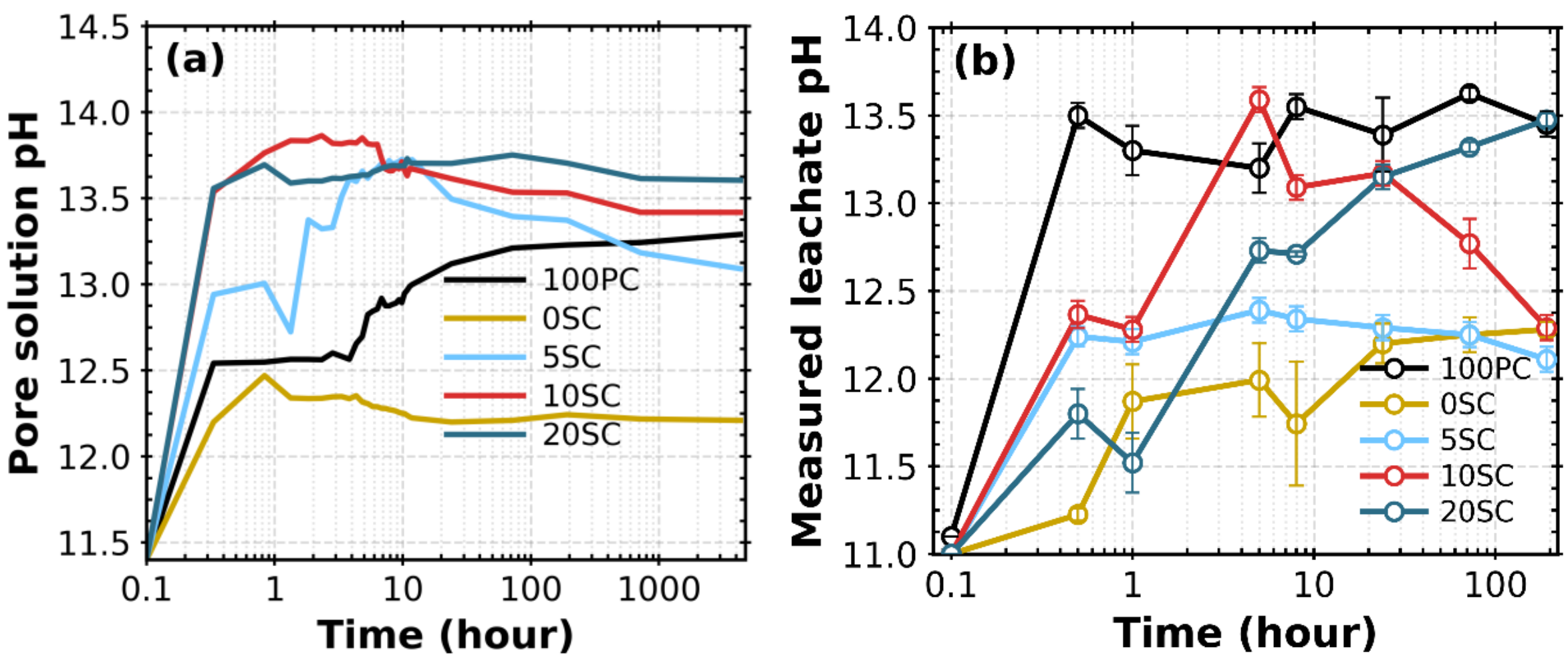

Figure 12. (a) Thermodynamically calculated evolution of pore solution pH and (b) experimentally measured pH evolution obtained from direct pore solution extraction at early ages and *ex situ* leaching at later ages for 100PC, 0SC, 5SC, 10SC, and 20SC. The later-age leachate values indicate relative alkalinity trends and do not represent undiluted *in situ* pore-solution pH.

Figure 13 presents the thermodynamically calculated temporal evolution of major dissolved aqueous species ($Ca^{2+}$, Si, Al, $Na^{+}$, $OH^{-}$, and $CO_3^{2-}$) for the four slag-cement systems. In 0SC, dissolved Ca associated with hydration of the PC fraction exceeds dissolved Si and Al by more than an order of magnitude, while slag dissolution remains limited under moderate alkalinity (pH ≈ 12.2–12.4). At ~24 hrs, Ca concentration decreases as it becomes incorporated into hydration products, accompanied by the onset of slag dissolution (Figure 8a). The overall evolution reflects a transition from PC-controlled to increasingly slag-influenced aqueous chemistry typical of high-volume slag systems.

In the $Na_2CO_3$-activated systems, the initial pore solution is characterized by elevated carbonate concentrations and higher alkalinity. In 5SC and 10SC, the early decrease in dissolved carbonate is accompanied by an increase in $OH^{-}$ concentration, consistent with redistribution of carbonate from the aqueous phase into carbonate-bearing solids and the associated development of a more alkaline pore solution. The resulting alkalinity promotes dissolution of slag, leading to pronounced increases in dissolved Si and Al species relative to 0SC. These species subsequently decrease as they are incorporated into the binder gel and other hydrates, broadly coinciding with the main heat-release peak observed in ICC. However, the persistence of alkalinity differs between these systems: in 5SC, $OH^-$ concentration declines relatively rapidly at later ages, corresponding to limited sustained slag dissolution and lower overall reaction extent, whereas in 10SC the more rapid and larger initial increase in alkalinity supports stronger slag activation and gel formation before the $OH^-$ concentration subsequently decreases.

The aqueous evolution of 20SC differs from that of the lower-dosage systems. Dissolved carbonate decreases substantially during 1–10 hrs, while the $OH^-$ concentration remains near its maximum. Alkalinity then declines only gradually over a substantially longer period. Thus, redistribution of carbonate into solid phases does not rapidly exhaust the alkaline environment in 20SC. This behavior is consistent with the larger total alkali inventory and the evolving partitioning of Na, Ca, carbonate, and Al between the pore solution and carbonate-bearing phases. The dissolved Si and Al concentrations in 20SC also reach markedly higher transient maxima than in 5SC and 10SC, particularly for Al. Their increase broadly coincides with the onset of the principal calorimetric reaction period, after which both concentrations decrease as gel formation proceeds. Aqueous concentrations reflect the balance between release through reactant dissolution and removal through hydrate precipitation. Combined with the QXRD evidence that substantial early slag reaction in 20SC is accompanied by comparatively limited contemporaneous gel accumulation, the elevated Si and Al concentrations support a transient regime in which release of aluminosilicate species outpaces their incorporation into C–(A)–S–H-type gel. The mechanistic implications of this dissolution–precipitation imbalance are developed in Section 3.6.

The experimentally measured aqueous ion concentrations also show qualitative consistency with the calculated dosage-dependent trends. Figure 14 compares the experimentally measured concentrations obtained by ICP-OES with the corresponding GEMS predictions and shows an overall positive correlation between the two datasets. The largest deviations are primarily associated with Ca concentrations in the activated systems, where the experimentally measured values were frequently near or below the ICP-OES detection limit. For Ca, the very low dissolved concentrations are consistent with rapid precipitation into carbonate-bearing phases and other hydration products. Despite differences in absolute concentrations, the overall positive correlations suggest that the calculated and measured results broadly reproduce the principal dosage-dependent trends and support the aqueous-chemistry interpretation rather than quantitatively validating the calculated concentrations.

Overall, the calculations and measurements identify distinct aqueous-chemistry regimes: moderate PC-derived alkalinity in 0SC; transient alkalinity enhancement in 5SC; rapid but less persistent activation in 10SC; and sustained alkalinity coupled with prolonged slag dissolution and transient accumulation of dissolved Si and Al in 20SC. These dosage-dependent regimes, along with the results in Sections 3.2–3.5, provide the chemical basis for the integrated reaction mechanism proposed in Section 3.6.

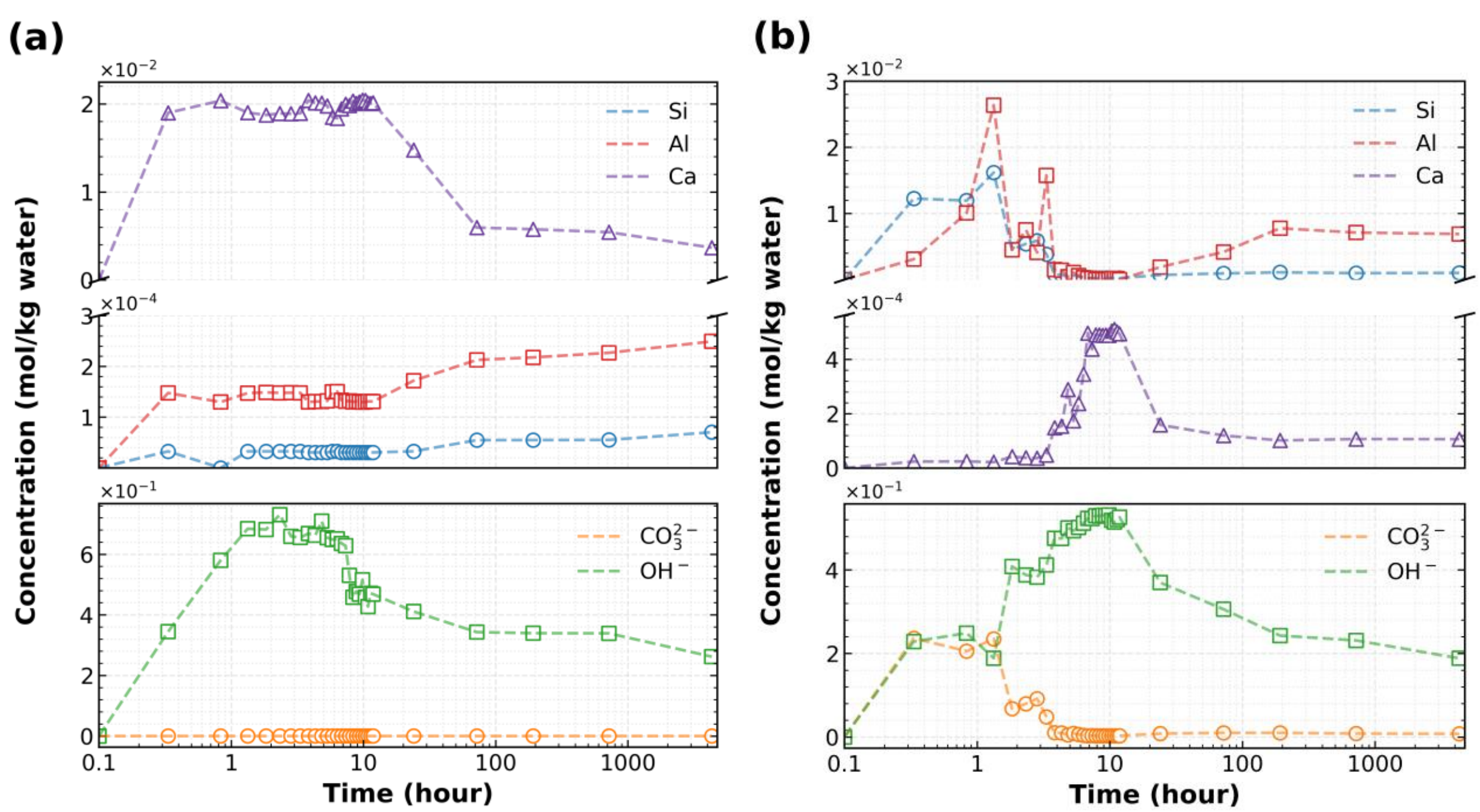

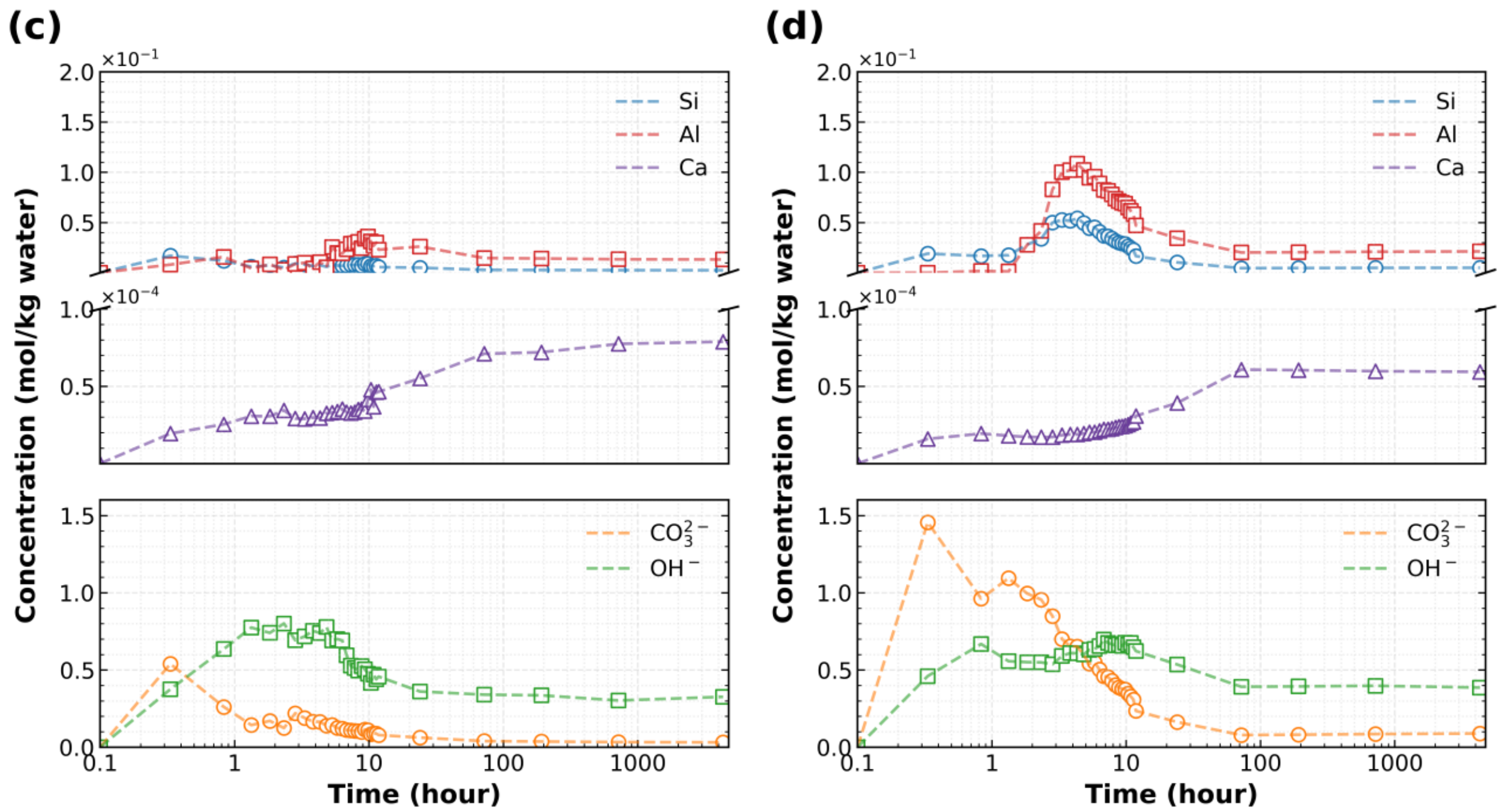

Figure 13. Thermodynamically calculated temporal evolution of total dissolved Ca, Si, Al, and Na and the principal OH⁻ and $CO_3^{2-}$ species for (a) 0SC, (b) 5SC, (c) 10SC, and (d) 20SC.

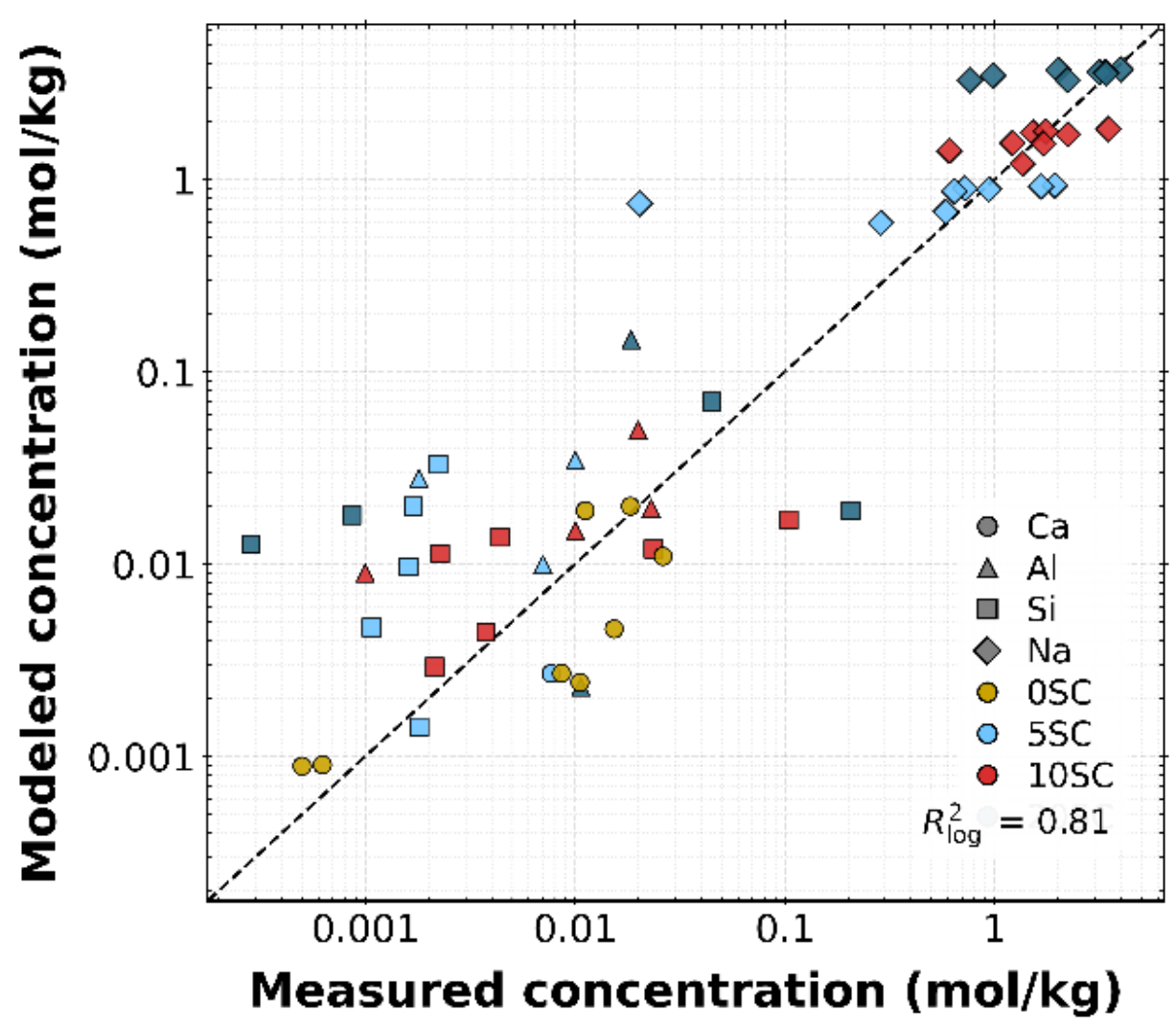

Figure 14. Comparison of thermodynamically calculated and experimentally measured aqueous concentrations obtained from directly extracted early-age pore solutions and later-age ex situ leachates by ICP–OES. The two extraction methods are distinguished, and the later-age leachate results are interpreted primarily as indicators of relative trends.

### 3.5.2 Porosity evolution and relationship with strength

Based on the thermodynamic outputs, total porosity was determined from the difference between initial volume before hydration and volume of solid-phase volume at each hydration age [139]. This pore volume was then partitioned into (i) gel porosity, representing nanoscale pore volume within the C–(A)–S–H-type gel and calculated using a fixed gel-pore fraction (0.30 for PC systems, 0.35 for alkali-activated systems), and (ii) capillary porosity, calculated as the residual pore volume after subtracting the estimated gel-pore contribution from the total pore volume [139–141]. More calculation details along with sensitivity analysis are given in Section S5.3 of Supporting Information. The resulting calculated evolution of total and capillary porosity is shown in Figure 15a. Among the four UHVS blends, 5SC exhibits the highest total and capillary porosity, whereas 20SC shows the lowest. The progressive reduction in total and capillary porosity, accompanied by an increase in gel porosity with increasing $Na_2CO_3$ dosage, reflects higher degrees of reaction and greater formation of solid phases, including the binder gel and carbonate- and AFm-type hydrates. These calculated porosity trends are consistent with the measured compressive strengths, with lower total and capillary porosity corresponding to higher mechanical performance, as reflected by the general inverse correlations shown in Figures 15b and 15c. Fitting of the data using Ryshkewitch–Duckworth exponential model [142,143] is shown in Figures 15b and 15c, with $R^2$ values of ~0.8. In particular, the higher later-age porosity of 5SC relative to 0SC is consistent with its lower 8- and 28-day strengths, whereas the lowest calculated porosity of 20SC (approaching that of 100PC) is consistent with its superior strength development over 0–10SC and comparable strength development with 100PC. Although these relationships do not independently validate the absolute calculated porosities, they provide a physically consistent link among reaction progress, pore-volume reduction, and strength development.

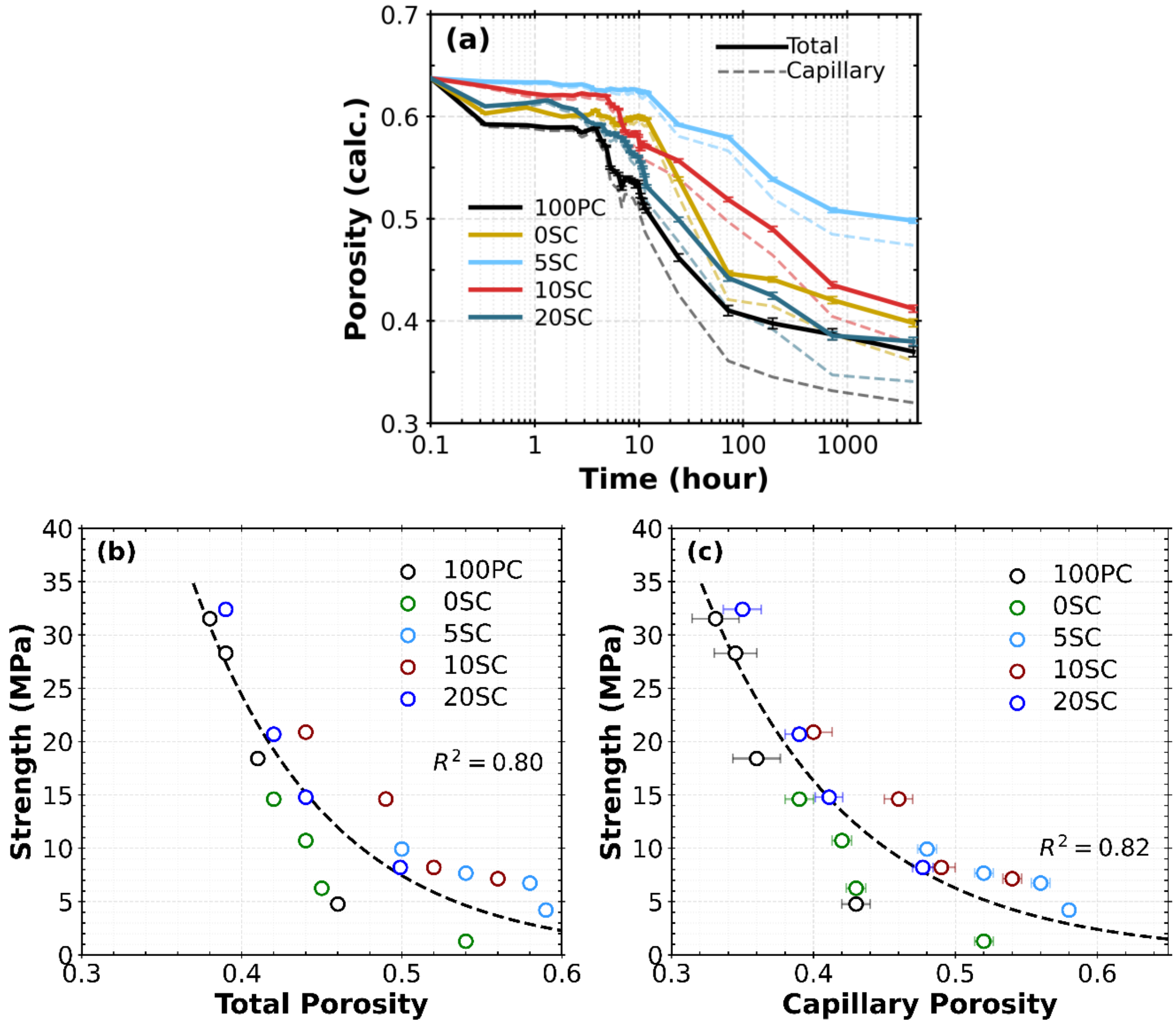

Figure 15. (a) Thermodynamically calculated evolution of total and capillary porosity for 100PC, 0SC, 5SC, 10SC, and 20SC. Error bars indicate the sensitivity of the calculated porosity to variation of the QXRD inputs within the TOPAS-reported fitting uncertainties and do not represent total model uncertainty. (b) total porosity and (c) capillary porosity for all modeled mixtures and curing ages. In (c)the scatter points represent the capillary porosity calculated using assumed gel-pore fraction 0.35, while the horizontal bars indicate the minimum-to-maximum range 0.20 to 0.40, illustrating the sensitivity of the strength–capillary porosity relationship to the assumed gel-pore fraction. Maximum relative change from baseline was about 4.93%, showing small effect on capillary porosity. Relationships between measured compressive strength and thermodynamically calculated and correlated using Ryshkewitch–Duckworth exponential model [142,143].

### 3.6 Reaction Mechanism

The combined calorimetric, QXRD, aqueous-chemistry, thermodynamic, and strength results demonstrate that $Na_2CO_3$ can effectively activate a UHVS containing 90 wt.% GGBS when its dosage is sufficient to sustain alkaline conditions throughout hydration. The activation process is governed not by the initial high pH alone, but by the time-dependent coupling among PC hydration, carbonate precipitation and redistribution, slag dissolution and hydrate formation. The proposed dosage-dependent mechanism supported by multi-technique evidence is summarized in Figure 16.

Immediately after mixing, $Na_2CO_3$ dissolves rapidly, releasing $Na^+$ and carbonate species, while dissolution and hydration of the PC fraction, particularly alite, supply $Ca^{2+}$ and contribute to alkalinity. Once the solution becomes supersaturated with respect to Ca–bearing carbonate phases, their precipitation removes dissolved Ca and carbonate, promotes continued PC reaction, and shifts the coupled phase-solution equilibria toward increased $OH^-$ availability [46,50]. In 5SC and 10SC, calcite dominates the early carbonate assemblage (Figures 8b and 8c), whereas the higher Na and carbonate activities in 20SC favor substantial early gaylussite formation (Figure 8d). The contrasting species partitioning can be represented as:

$$\mathrm{Na_2CO_3}(aq) + \mathrm{Ca(OH)_2^*} \rightarrow \mathrm{CaCO_3(s)} + \mathrm{2NaOH(aq)}$$

$$\mathrm{2Na_2CO_3}(aq) + \mathrm{Ca(OH)_2^*} + \mathrm{5H_2O} \rightarrow \mathrm{Na_2Ca(CO_3)_2} \cdot \mathrm{5H_2O(s)} + \mathrm{2NaOH(aq)}$$

Here, $\mathrm{Ca(OH)_2^*}$ represents the Ca–alkalinity equivalent supplied through PC hydration rather than experimentally detected crystalline portlandite, and the equations are intended as species-partitioning balances rather than formal net ionic reactions. Calcite incorporates $Ca^{2+}$ and carbonate but no $Na^+$, whereas gaylussite incorporates two $Na^+$ ions for every two carbonate groups. On this simplified basis, the gaylussite-forming pathway temporarily retains a larger fraction of activator-derived $Na^+$ in the solid and nominally leaves approximately half as much alkalinity in solution per mole of $Na_2CO_3$ as the calcite-forming pathway. This provides a plausible explanation for the lower initial pH of 20SC relative to 10SC despite its higher $Na_2CO_3$ dosage.

In 0SC, hydration of the 10 wt.% PC fraction produces only moderate alkalinity, resulting in comparatively slow slag reaction. Appreciable slag consumption and binder gel formation are first detected at ~24 hrs (Figure 8a), coinciding with the principal calorimetric response (Figure 4). In 5SC, the increase in alkalinity accelerates early reaction (Figure 4a) and produces higher 1- and 3-day strengths than 0SC (Figure 3). However, the limited alkali inventory is insufficient to sustain later slag reaction, and gel accumulation falls behind that of 0SC (Figure 8b). This results in higher calculated porosity (Figure 15) and lower 8- and 28-day strengths (Figure 3). Thus, 5SC represents an early but under-dosed activation regime.

Increasing the dosage to 10SC produces rapid and strong early activation. Calcite-dominated precipitation leaves most activator-derived Na in solution and coincides with the fastest and highest initial pH increase among the investigated mixtures (Figures 8c and 12). The resulting alkalinity substantially accelerates reactions, promoting closely coupled slag dissolution and C–(A)–S–H-type gel formation, with the gel content reaching ~20% by ~12 hrs (Figure 8c). The pH subsequently declines, however, falling below that of 20SC at ~12 hrs in the thermodynamic calculations and by ~24 hrs in the experimental measurements (Figure 12). This rapid early binder gel precipitation may increase reaction-product coverage of slag surfaces and restrict subsequent mass transport, leading to lower reaction extent and coarse pore structure at later ages, as has been reported for rapidly activated slag systems [144,145], although this was not directly resolved here.

The 20SC system follows a different pathway. Gaylussite formation initially stores part of its larger Na and carbonate inventory in a transient solid phase, suppressing the immediate pH increase relative to 10SC (Figure 12). As hydration proceeds, however, gradual gaylussite re-equilibration may redistribute $Na^{+}$, $Ca^{2+}$, and carbonate between the solid and aqueous phases. Released $Ca^{2+}$ and carbonate may subsequently be redistributed into Na-free carbonate-bearing phases (Hc, Mc, and calcite), while released $Ca^{2+}$ may also contribute to evolving C–(N)–A–S–H-type gel assemblage; part of the released $Na^{+}$ may remain aqueous. Dissolved carbonate can additionally contribute to alkalinity through carbonate hydrolysis:

$$CO_3^{2-} + H_2O \rightleftharpoons HCO_3^{-} + OH^{-}$$

Gaylussite therefore may not directly generate or conventionally buffer $OH^{-}$. Rather, it may act as a transient alkali-distribution reservoir that delays the aqueous availability of part of the larger $Na_2CO_3$ inventory. Together with the greater total alkali content of 20SC, this time-dependent redistribution provides a plausible explanation for its lower initial but more persistent pH relative to 10SC (Figure 12).

During the first ~12 hrs, slag dissolution in 20SC substantially outpaces binder gel precipitation (Figure 8d). By ~12 hrs, ~ 22% of the slag has reacted, compared with ~14% in 10SC, yet the corresponding binder gel contents are only ~8% compared with ~20% in 10SC (Figures 8c–d). Instead, gaylussite and Hc reach ~10% and ~7%, while the thermodynamic calculations indicate markedly higher transient dissolved Si and Al concentrations, particularly Al (Figure 13). Although the modeled aqueous concentrations remain uncertain because the model overpredicts early gel formation for 20SC (Figure 11d), the combined results indicate a temporary decoupling between slag dissolution and incorporation of the released species into the principal binding gel. Preferential Ca partitioning into gaylussite and carbonate-AFm phases may limit immediate C–(A)–S–H-type gel precipitation, while the persistent alkalinity continues to promote slag dissolution.

Between 12 and 24 hrs, 20SC transitions from this carbonate-rich assemblage toward rapid gel formation (Figure 8d). Residual slag decreases by ~11 wt.%, gaylussite decreases from ~10% to ~7%, and Hc decreases from ~7% to ~4%, while the QXRD-assigned binder gel content increases from ~8% to ~25%. These decreases are not accompanied by measurable calcite growth, and Mc remains below ~1%. The concurrent changes therefore support a combination of continued slag dissolution, consumption of the transient dissolved Si–Al reservoir, and possible re-equilibration of the early carbonate-bearing phases. The numerical correspondence between the decreases in slag and transient phases and the increase in gel should not be interpreted as direct stoichiometric conversion because the phases differ in composition, the gel incorporates chemically bound water, and aqueous or poorly resolved products are not fully represented by QXRD.

Beyond 24 hrs, continued redistribution among carbonate-bearing phases is evident. From 24 hrs to 8 days, Mc increases from <1% to ~7% as gaylussite gradually decreases, while Hc and calcite remain near ~4% and ~3%, respectively. This evolution is consistent with progressive partitioning of carbonate into Na-free AFm phases, potentially increasing the fraction of activator-derived $Na^+$ retained in solution. Between 8 and 28 days, Hc and Mc become undetectable, while calcite increases from ~3% to ~7%. Over the same interval, residual slag decreases from ~40% to ~31%, whereas the QXRD-assigned gel content increases from ~35% to ~54%. These changes suggest continued slag dissolution and progressive re-equilibration of the carbonate assemblage: Ca and Al released from the disappearing AFm phases may contribute to the evolving gel assemblage, while part of their carbonate may be redistributed into calcite. The larger gel increase relative to slag consumption is also consistent with incorporation of chemically bound water and redistribution of elements from pre-existing hydrates, rather than gel formation solely from newly reacted slag.

At later ages, gaylussite persists but gradually decreases to below ~5% by 180 days, while calcite increases further to ~9%. These trends are consistent with progressive redistribution of carbonate from Na-bearing gaylussite and transient AFm phases toward more stable Na-free carbonate-bearing phases. Such redistribution may help maintain a larger dissolved alkali inventory and thereby, through coupled aqueous speciation, contribute to the persistent alkalinity of 20SC. Together with the larger initial $Na_2CO_3$ inventory, this evolving phase–solution partitioning supports continued slag dissolution and progressive gel formation. At 180 days, the residual slag and gel contents reach ~28% and ~58%, respectively (Figure 8d), consistent with the higher cumulative heat release (Figure 4b), lowest calculated total and capillary porosities (Figure 15), and superior strength among the slag-cement mixtures studied (Figure 3).

Overall, increasing $Na_2CO_3$ dosage shifts the binder through three activation regimes: early but under-dosed activation in 5SC, rapid but less persistent activation in 10SC, and sustained activation with temporary dissolution–precipitation decoupling and prolonged carbonate redistribution in 20SC. Effective activation therefore requires sufficient alkali inventory and evolving phase-solution partitioning that maintain favorable solution conditions as hydration proceeds. This framework explains the non-monotonic influence of $Na_2CO_3$ dosage on reaction kinetics, phase evolution, gel formation, calculated porosity, and compressive strength.

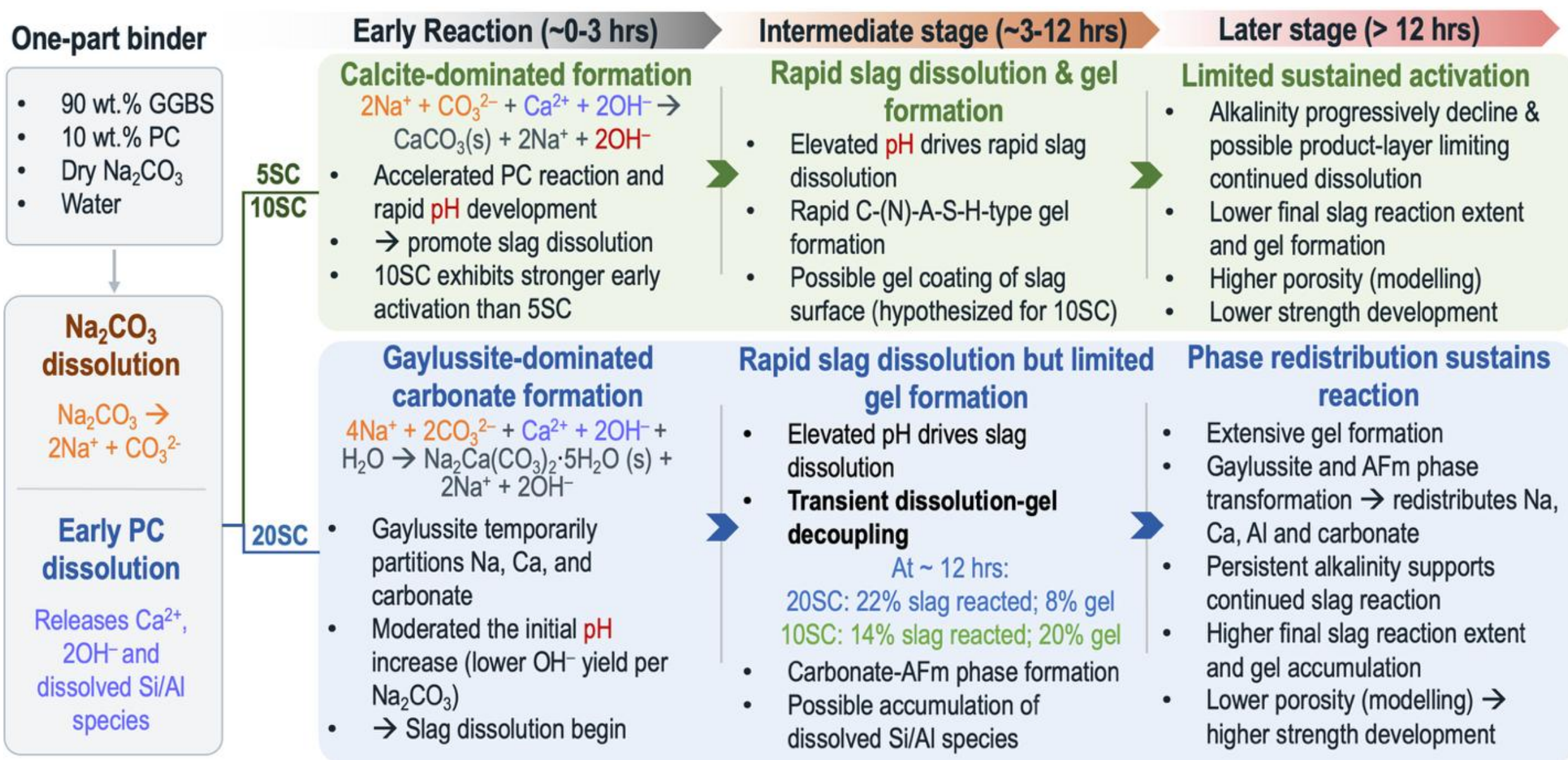

Figure 16. Proposed dosage-dependent reaction pathways in $Na_2CO_3$-activated 90% GGBS–10% PC binders. At low-to-intermediate $Na_2CO_3$ dosages (5SC and 10SC), calcite-dominated carbonate reactions promote rapid early alkalinity development and slag activation, but declining alkalinity limits sustained reaction. In contrast, 20SC exhibits substantial early formation of gaylussite and carbonate-AFm phases, followed by transient dissolution–gel decoupling and subsequent phase–solution redistribution associated with more persistent alkalinity, prolonged slag reaction, and greater later-age gel formation.

### 3.7 Screening-level Life Cycle Assessment and Material-cost Implications

To provide an initial indication of the environmental and economic implications of the mechanically viable 20SC formulation, a screening-level cradle-to-gate global warming potential (GWP) and material-cost assessment was conducted for several $Na_2CO_3$ production pathways (Figure 17). The analysis evaluates whether the relatively high $Na_2CO_3$ dosage offsets the environmental benefit of replacing 90 wt.% of PC with GGBS and identifies the upstream processes controlling the results. It is not intended as a comprehensive application-level sustainability assessment because fresh-state behavior, dimensional stability, durability, service life, scale-up, and other practical performance requirements were not evaluated.

The 100PC reference system (S0) exhibits the highest cradle-to-gate GWP, approximately 0.85 kg $CO_2$-eq $kg^{-1}$ binder, because of emissions associated with limestone calcination and kiln energy demand. This value is consistent with commonly reported cradle-to-gate emission intensities for Portland cement [146–149]. Replacing 90 wt.% of PC with GGBS without $Na_2CO_3$ activation (S1) reduces the calculated GWP by about 80%, primarily because of the lower assigned emission intensity of GGBS (~0.1 kg $CO_2$-eq $kg^{-1}$). However, this formulation exhibits substantially delayed strength development (Figure 3). When activated by 20 wt.% $Na_2CO_3$ (20SC), the strength development is restored to a level approaching that of 100PC (Figure 3), while increasing the calculated GWP to ~0.25–0.31 kg $CO_2$-eq $kg^{-1}$ binder because of the upstream burden of $Na_2CO_3$ production. Nevertheless, this still represents an approximately 66–72% reduction relative to 100PC, depending on the $Na_2CO_3$ production pathway.

The evaluated $Na_2CO_3$ pathways include recovery of natural soda ash from trona ore (S2), synthetic production via the Solvay process (S3), and an emerging $CO_2$ capture-integrated route involving chlor-alkali-derived NaOH followed by carbonation with captured $CO_2$. The latter route was evaluated using both fossil-dominated (S4) and renewable-intensive (S5) electricity mixes. Natural soda ash gives the lowest calculated GWP among the evaluated pathways, whereas the $CO_2$ capture-integrated route is strongly affected by the electricity intensity of upstream NaOH production. These results suggest that incorporation of captured $CO_2$ does not inherently ensure a low-carbon $Na_2CO_3$ supply pathway.

The preliminary material-cost assessment indicates that replacing 90 wt.% of PC with GGBS alone does not reduce material cost under the adopted price assumptions because the unit costs assigned to GGBS and PC are similar. Addition of $Na_2CO_3$ increases the calculated material cost by approximately 9–23% relative to 100PC, depending on the production pathway. Natural soda ash provides the lowest cost among the evaluated $Na_2CO_3$ routes, whereas the $CO_2$ capture-integrated scenarios remain more expensive because of the costs associated with NaOH production, electricity, $CO_2$ capture, and downstream processing.

Overall, the screening-level analysis indicates that the relatively high $Na_2CO_3$ dosage does not eliminate the substantial cradle-to-gate GWP advantage created by replacing 90 wt.% of PC with GGBS, although it may increase material cost under current price assumptions. Carbon pricing could improve the relative economic competitiveness of the binder by assigning value to its avoided emissions, but this possibility was not quantified in the present analysis. Future mix-design optimization, partial use of lower-cost supplementary cementitious materials, and development of lower-impact $Na_2CO_3$ production routes may further improve both environmental and economic performance. These results demonstrate the feasibility of $Na_2CO_3$ activation as a potential pathway toward low-carbon UHVS systems.

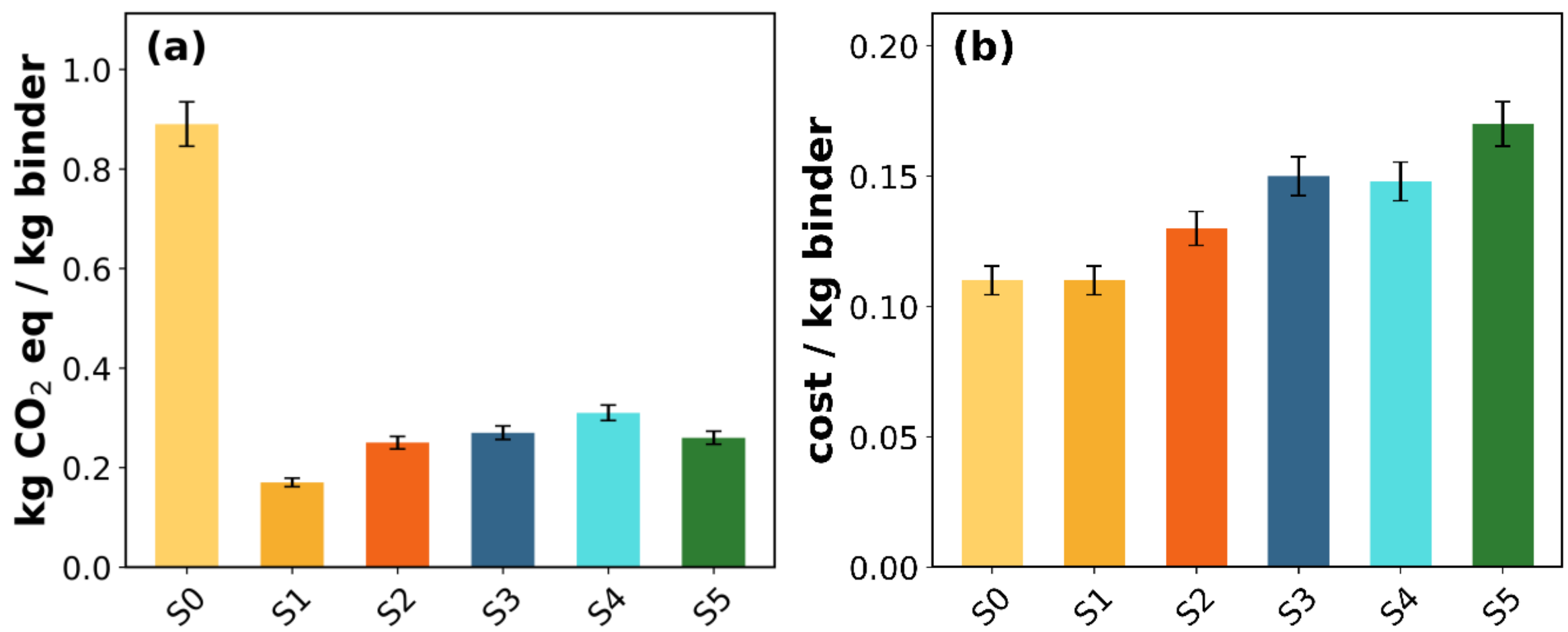

**Figure 17.** Screening-level cradle-to-gate GWP and material-cost assessment for all binder production scenarios (S0–S5). **(a)** Global Warming Potential (GWP; kg $CO_2$-eq per kg binder) and **(b)** material cost (USD per kg binder) for the reference Portland cement system (S0: 100PC), unactivated UHVS system (S1: 0SC), and those activated by 20% $Na_2CO_3$ (S2-S5: 20SC). S2-S5 represent different $Na_2CO_3$ production and sourcing pathways. The error bars are obtained via Monte Carlo simulation (details given in Section 6 of Supporting Information).

## 3.8 Broader implications, limitations and future work

### *3.8.1 Broader implications*

This work demonstrates that tuning the $Na_2CO_3$ dosage provides a promising pathway to activate a UHVS binder containing 90 wt.% GGBS and achieve compressive strengths approaching the PC reference. When appropriately dosed, $Na_2CO_3$ enables sustained slag dissolution and gel formation without relying on highly caustic hydroxide- or silicate-based activators. The one-part system, in which $Na_2CO_3$ is dry-blended with the binder and only water is added during mixing, can simplify handling relative to conventional two-part alkali-activated systems [150]. The screening-level cradle-to-gate life-cycle and material cost assessments indicate that these binders can achieve substantially lower $CO_2$ emissions than the PC reference, albeit with a moderate cost premium under the adopted inventory and market assumptions. This cost premium could be mitigated under carbon-pricing policies or through partial replacement of slag with lower-cost supplementary cementitious materials (e.g., fly ash) or fillers (e.g., limestone powder).

Beyond the specific binder system studied, this work illustrates the value of combining ICC, time-resolved QXRD, aqueous-chemistry measurements, FTIR spectroscopy, and experimentally constrained thermodynamic modeling to resolve reaction kinetics and mechanisms in chemically complex cementitious systems. The integration of time-resolved experiments with thermodynamic constraints enables an internally consistent interpretation of heat release, phase assemblage, pore-solution chemistry, precursor reaction, gel formation, calculated porosity, and mechanical performance. This coupled methodology provides a potentially transferable framework for other alkali-activated and blended cement systems, supporting mechanistic understanding and composition–reaction–property-based materials design.

### *3.8.2 Limitations and future work*

Several aspects of the present study define the scope of the conclusions and motivate further investigation. First, the investigation focused on a single source of GGBS and PC at a fixed blend ratio (90 wt.% GGBS–10 wt.% PC) and water-to-binder ratio. Given the known variability in slag chemistry, glass content, fineness, and sulfur content and speciation [108,151], the optimal $Na_2CO_3$ dosage and the balance between carbonate formation and gel development may shift for other precursor combinations. Extending this framework to a broader range of slags, cement contents, water-to-binder ratios, and curing conditions is therefore necessary to assess the generality of the identified reaction pathway. Additionally, the 90 wt.% GGBS formulation investigated here is used as a model UHVS. Future studies could examine whether comparable activation and strength can be achieved at lower GGBS contents through multicomponent formulations incorporating other SCMs, such as fly ash or metakaolin, and fillers, such as limestone. Such formulations could reduce dependence on GGBS supply and potentially lower material costs while retaining the benefits of one-part $Na_2CO_3$ activation. Because these substitutions would alter Ca, Al, Si, and carbonate availability, the $Na_2CO_3$ dosage and phase–solution balance would need to be optimized.

Second, because the study focused on elucidating reaction kinetics and mechanisms, the performance evaluation was limited to mortar compressive strength. Future work should also investigate the fresh-state properties (e.g., workability, rheology, and setting behavior), pore-structure development, as well as long-term dimensional stability and durability-related properties of these systems. Shrinkage warrants particular attention given its known importance in alkali-activated slag materials [40,41,45].

Third, both the quantitative phase analysis and thermodynamic modeling involve inherent methodological limitations. The QXRD-assigned C–(A)–S–H-type gel was represented using a PONKCS pseudo-phase and therefore does not directly resolve gel composition or nanoscale structural heterogeneity. The thermodynamic calculations were constrained using experimentally determined reaction extents and assumed equilibrium or local-equilibrium conditions. As a result, kinetic barriers, transport limitations, and spatial heterogeneity were not explicitly captured, and metastable phases such as gaylussite required imposed constraints rather than being predicted independently. Although the sensitivity analyses presented here indicate that these assumptions do not alter the principal mechanistic conclusions, future efforts combining kinetic or reactive transport modeling with time-resolved experiments would provide a more complete description of the evolving phase-solution dynamics.

Fourth, the environmental and economic assessments are screening-level evaluations based on a binder-mass functional unit. The calculated outcomes remain sensitive to assumptions regarding GGBS allocation, $Na_2CO_3$ production routes, electricity sources, transportation, and material prices. The assessments also do not account for differences in fresh-state performance, construction requirements, durability, service life, or scale-up. If subsequent studies further reinforce the technical viability of this binder concept, more comprehensive LCA and technoeconomic analysis would be worthwhile.

Finally, while the global supply of GGBS is limited, slag remains one of the most widely used low-carbon SCMs, and improving its utilization efficiency offers immediate decarbonization benefits. More broadly, the reaction pathways identified in this study suggest a potentially transferable activator-design principle: effective carbonate activation can be promoted by sustaining alkalinity through time-dependent phase–solution partitioning rather than simply maximizing initial pH. In this sense, the system investigated here serves as a model for developing broader principles of mild-alkali activation and SCM-rich binder design.

# 4 CONCLUSION

UHVS binders can substantially reduce clinker ratio, but their application is limited by slow slag reaction and delayed strength development. Conversely, $Na_2CO_3$ offers a less caustic and potentially lower-carbon alternative to NaOH- and silicate-based activators for alkali-activated materials, yet its moderate intrinsic alkalinity often leads to similarly delayed reactions and strength development. Integrating $Na_2CO_3$ activation with a UHVS binder therefore offers a potential pathway to mitigate the limitations of both systems. Here, this concept was demonstrated in a binder containing 90 wt.% GGBS and 10 wt.% PC: by tuning the $Na_2CO_3$ dosage, the 20SC formulation (20% $Na_2CO_3$) achieved sustained slag activation, and its 28-day compressive strength approached that of the 100PC reference under the tested conditions.

$Na_2CO_3$ dosage fundamentally altered both the rate and persistence of activation by controlling the formation of Ca-bearing carbonate phases and the temporal evolution of alkalinity. Addition of 5% $Na_2CO_3$ (5SC) accelerated early hydration and strength development, but this acceleration was not sustained, and later slag reaction and gel accumulation fell below those of 0SC. Increasing the dosage to 10% (10SC) produced the highest initial pH, rapid slag dissolution and gel formation, and enhanced early strength. However, the subsequent decline in alkalinity resulted in a less persistent reaction pathway and limited later-age development.

In contrast, 20SC followed a staged activation pathway characterized by lower initial but more persistent alkalinity. Its higher Na and carbonate activities favored substantial early formation of gaylussite and carbonate-AFm phases, which temporarily partitioned part of the activator-derived Na and carbonate into the solid assemblage. Together with the larger total alkali inventory, subsequent phase–solution re-equilibration helped maintain favorable alkaline conditions for continued slag dissolution and progressive gel formation. This sustained activation pathway resulted in the highest slag reaction degree and later-age binder gel content, along with the lowest calculated total and capillary porosities, leading to strength development approaching that of the PC reference. The results therefore demonstrate that effective $Na_2CO_3$ activation is governed not by the maximum initial pH, but by the persistence of favorable alkaline conditions controlled through dosage-dependent carbonate chemistry and phase–solution partitioning.

The screening-level cradle-to-gate assessment further indicates that the relatively high $Na_2CO_3$ dosage does not eliminate the potential carbon benefit of replacing 90 wt.% of PC with GGBS. Depending on the $Na_2CO_3$ supply pathway, the 20SC formulation provides an estimated 66–72% reduction in GWP relative to 100PC, although its calculated material cost is approximately 9–23% higher under the adopted assumptions. Carbon pricing and future mix-design optimization may improve the relative economic competitiveness of the system, but these possibilities require further evaluation.

Overall, this work establishes a mechanistically grounded framework for one-part, mild-carbonate activation of UHVS binders. Although validation across different precursor sources and evaluation of fresh-state behavior, dimensional stability, and durability remain important, the reaction pathway identified here provides a potentially transferable design principle for mild-alkali activation: effective carbonate activation depends on sustaining favorable alkalinity through time-dependent phase–solution partitioning.

## 5 APPENDIX A. SUPPORTING INFORMATION

The supporting information contains: **Section 1. Estimation of degree of reaction from isothermal conduction calorimetry (ICC) and quantitative XRD (QXRD)**, covering the calculation, comparison and sensitivity analysis of reaction extents derived from calorimetry and QXRD; **Section 2. *In Situ* X-ray diffraction sample configuration**, describing the sample-holder configuration and measurement setup used for *in situ* XRD analysis; **Section 3. Detailed methodology of QXRD analysis**, covering refinement of the corundum internal standard, Kapton-film and free-water pseudo-PONKCS phases, Rietveld refinement of PC powder and comparison with XRF oxide composition, development of PONKCS models for the amorphous GGBS fraction and C–(A)–S–H, and refinement procedures for *in situ* samples; **Section 4. Empirical interpolation of GGBS, alite, belite and gaylussite evolution**, describing the fitting and interpolation procedures used to obtain continuous phase-evolution profiles; **Section 5. Thermodynamic phase evolution calculated via GEMS**, covering the thermodynamic modeling procedure, phase-evolution calculations, gel- and capillary-porosity estimation along with sensitivity analysis; **Section 6. Life cycle assessment and cost analysis**, covering goal and scope definition, scenario definitions, modeling assumptions and boundary conditions, system boundaries, life-cycle inventories for GGBS and alternative $Na_2CO_3$ production pathways, life-cycle impact assessment, and preliminary cost analysis; **Section 7. Initial calorimetric peak and correlation between strength and cumulative heat**, providing supplementary calorimetric comparisons and strength–heat relationships; **Section 8. Full FTIR spectrum for all the mixes**; **Section 9. Comparison of experimentally measured and thermodynamically predicted pH**; and **Section 10. Minor phases quantified in QXRD for the mixes**.

## 6 DECLARATION OF GENERATIVE AI USE

During the preparation of this work, the authors used OpenAI ChatGPT to assist with language refinement, coding, and reference and citation checking. All AI-assisted outputs and literature-related suggestions were critically reviewed, independently verified, and edited by the authors. The authors take full responsibility for its final content.

## 7 DECLARATION OF COMPETING INTEREST

The authors declare no competing interests.

## 8 ACKNOWLEDGMENTS

This work was supported by the Department of Civil and Environmental Engineering at Rice University and the Gulf Research Program's Early-Career Research Fellowship. Z.L.'s participation is partially supported by the Rice Academy Postdoctoral Fellowship. The authors would like to acknowledge Dr. Jianhua Li and Dr. Gang Liang for their technical support and valuable advice on using the equipment hosted by the Shared Equipment Authority (SEA) at Rice University. The authors also acknowledge Sesco Cement Corporation, Houston, Texas, USA, for providing raw Portland cement used in this study.